\documentclass[aps,floats,pre,showpacs,twocolumn
]{revtex4}

\newcommand{\hL}{{\hat {\cal L}}'_{12}}
\newcommand{\V}{{\bf  V}}
\newcommand{\C}{{\bf  C}}

\newcommand{\G}{{\bf \Gamma}}
\newcommand{\E}{{\hat {\cal E}}}
\newcommand{\U}{{\hat {\frak U}}}
\newcommand{\Li}{{\hat {\frak L}}}
\newcommand{\GG}{{\hat {\frak g}}}
\newcommand{\R}{{\hat {\frak R}}}
\newcommand{\LI}{{\hat {\cal L}}}
\newcommand{\Y}{{\hat {\cal U}}}
\newcommand{\Z}{{\hat Z}}

\usepackage{amsfonts,amsmath} \usepackage{bm} \usepackage{dcolumn}
\usepackage{epsfig} \usepackage{latexsym}

\begin{document}

\title{Manifestly covariant classical correlation dynamics II. \\
Transport equations and Hakim equilibrium conjecture}

\author{Chushun Tian}
\affiliation{Institut f{\"u}r Theoretische Physik, Z{\"u}lpicher
Str. 77, K{\"o}ln, D-50937, Germany}

\date{\today}
\begin{abstract}
 {\rm This is the second of a series of papers on the special relativistic
 classical statistical mechanics.
 Employing the general theory developed in the first paper we rigorously derive
 the relativistic Vlasov, Landau and Boltzmann equation. The latter two
 advocate the J{\"u}ttner distribution as the
 equilibrium distribution. We thus, at the full microscopic level,
 provide a support for the recent numerical finding [D. Cubero {\it et. al.},
 Phys. Rev. Lett. \textbf{99}, 170601 (2007)] of the special relativistic generalization
 of the Maxwell-Boltzmann distribution. Furthermore, the present theory allows us to rigorously
 calculate various correlation functions at the relativistic many-body equilibrium. Therefore, the
 relativistic many-body equilibrium conjecture of Hakim is justified.}
\end{abstract}

\pacs{03.30.+p, 52.25.Dg}
\maketitle

\section{Introduction}
\label{introduction}

The kinetic theory is one of the pillars of studies of relativistic
transport phenomena in various systems ranging from star clusters or
galaxies \cite{Kremer07,Israel84,Kandrup84,Kandrup86} to plasmas in
fusion \cite{Lefebvre08}, quantum chromodynamics
\cite{Heinz83,Litim02} and graphene discovered very recently
\cite{graphene}. The manifestly covariant counterparts of various
classical transport equations, such as the kinetic equation of
Vlasov, Fokker-Planck, Landau and Boltzmann, were proposed long time
ago (for a review see, for example, Ref.~\cite{Hakim67}), and their
applications nowadays have been well documented
\cite{Cercigani02,vanLeeuwen80}. These manifestly covariant
transport equations have received justifications from various
microscopic approaches
\cite{Israel84,Kandrup84,Hakim67,Prigogine65,Klimontovich60,DuBois72,Hu88}.
It is rather typical in rederiving relativistic transport equations
(for examples, see \cite{Hakim67,Klimontovich60,Hu88}) that the
Liouvillian dynamics of complete many-body distribution function is
bypassed and, crucially, the truncation approximation is resorted
to. Therefore, despite of the great success of this kind of
microscopic approaches \cite{Hakim67,Klimontovich60,Hu88} a
fundamentally important problem remains unsolved. That is, {\it will
a manifestly covariant transport equation be compatible with the
Liouvillian dynamics of the complete many-body distribution
function?} This issue has been addressed continuously for several
decades by many workers
\cite{Israel84,Kandrup84,Hakim67,Prigogine65,Balescu64,Schieve89,Lu94,Yaacov95}
and been far less advanced. This is by no means of pure theoretical
interests, rather, may find considerable practical applications.
Indeed, experiences in the Newtonian physics have shown that to go
beyond the weak coupling and Markovian approximation is inevitable
in order to understand collective dielectric effects in
electromagnetic plasmas \cite{Balescu60} and the infrared divergence
of the Fokker-Planck equation describing the stellar dynamics
\cite{Prigogine66,Kukharenko94}. In an insightful paper Kandrup
first realized that a Liouville equation of the complete many-body
distribution function admits an exact closed (nonlinear) kinetic
equation which is satisfied by the reduced one-body distribution
function and manifestly covariant \cite{Kandrup84}. Various special
transport equations, remarkably, may be unified within this general
kinetic equation. Unfortunately, he encountered principal
difficulties when proceeded further to derive the general kinetic
equation explicitly. Such a big gap is filled in the first of this
series of papers (denoted as Paper I) by the developed manifestly
covariant classical correlation dynamics \cite{Tian09}. One of the
main subjects of the present paper is to recover various special
transport equations systematically from the exact, manifestly
covariant and closed kinetic equation \cite{kinetic} found in Paper
I, which we call the {\it general transport equation}.

The existence of the general transport equation has important
implications on several aspects of classical special relativistic
(nonequilibrium) statistical mechanics which are undergoing
intensified studies. First of all, the mathematical structure
possessed by the general transport equation is in excellent
agreement with various approximate transport equations obtained by
other microscopic approaches
\cite{Israel84,Kandrup84,Prigogine65,Hu88,Klimontovich60,DuBois72,Lu94,Chou85},
but remarkably different from that proposed by Horwitz and coworkers
\cite{Schieve89}. Recently triggered by the latter there have been
many scientific activities \cite{Hanggi07,Debbasch08} searching for
the special relativistic generalization of the Maxwell-Boltzmann
distribution. The general transport equation reinforces the concept
well established by approximate transport equations. That is, the
J{\"u}ttner distribution serves as the relativistic (one-body)
equilibrium in dilute systems. Therefore, it is suggested that other
alternatives to the J{\"u}ttner equilibrium \cite{Schieve89} might
be specific to the (deterministic) relativistic many-body dynamics.

Secondly, in past years the relativistic Brownian motion and
diffusion have experienced considerable conceptual developments and
found important practical applications \cite{Hanggi05}. Long time
ago it was known that in the Minkowski spacetime nontrivial Lorentz
invariant Markovian processes do {\it not} exist
\cite{Dudley65,Hakim68}. It turns out that the relativistic Brownian
motion is interpreted as relativistic Markovian processes in the
$\mu$ phase space, and the latter is completely described by a
relativistic Fokker-Planck-type equation. So far these observations
have been investigated thoroughly at the level of one-particle
physics \cite{Hanggi05}, and Paper I is the first to substantiate
these important observations at the level of the {\it genuine}
relativistic many-body physics. Indeed, the general transport
equation arises from (i) the thermodynamic limit namely the particle
number $N\rightarrow +\infty$ and, (ii) that at given (global)
proper time particles lose the memory of the history of the entire
system. The condition (ii) is in sharp contrast to the deterministic
relativistic many-body dynamics, where the particle interaction is
profoundly nonlocal in spacetime, and serves as the many-body
dynamical origin of the Markovian processes in the $\mu$ phase
space. In particular, in Paper I the proper time parametrized
equation in the $8$-dimensional $\mu$ phase space is rigorously
justified, which was first obtained by Hakim \cite{Hakim68}, and
important roles of which have very recently been reinforced
\cite{Hanggi09}. From these perspectives the theory presented in
this series of papers may be considered as a microscopic approach to
(classical) special relativistic Brownian motion complementary to
the one based on the relativistic Langevin equation. In particular,
it might be proven to be a useful technique in exploring the
concepts such as relativistic noises, friction and fluctuation
theorem.

There is an adjacent important yet unsolved problem that may be
explored in the present theoretical scope. That is, to formulate the
relativistic many-body equilibrium. In the notable critical analysis
\cite{Hakim67} Hakim conjectured that at equilibrium there might
exist an infinite Lorentz invariant hierarchy of correlation
functions which is invariant under the spacetime translation and is
merely determined by the J{\"u}ttner distribution. Unfortunately,
further progresses have been impeded by the truncation approximation
intrinsic to various microscopic approaches (for examples,
Refs.~\cite{Hakim67,Hu88}), and the (dis)proof so far has been
missing. The general principles given in Paper I paves the way
towards justifying this conjecture. There the hierarchy of
(physical) correlation functions is found explicitly, which is
merely determined by the (physical) one-body distribution function.
Provided that the general transport equation admits an equilibrium
distribution, then the hierarchy of equilibrium correlation
functions is uniquely determined. To carry out this program for
rarified electromagnetic plasmas constitutes another main subject of
the present paper.

The present paper is written in the self-contained manner. The
readers that would not like to access to the mathematical foundation
may skip Paper I. The paper is organized as follows:
Sec.~\ref{generalformalism} is an exposition of the main results of
Paper I, and the exact starting point of the present paper is
pointed out. The rest is devoted to applications in relativistic
plasmas with electromagnetic interactions. In Sec.~\ref{transport}
we derive the relativistic Vlasov, Landau and Boltzmann equation
from the general transport equation found in Paper I. In
Sec.~\ref{equilibrium} we show that the present theory fully agrees
with the recent numerical finding \cite{Hanggi07} and advocates the
J{\"u}ttner distribution as the special relativistic generalization
of the Maxwell-Boltzmann distribution. The Hakim equilibrium
conjecture is justified, and the two-body equilibrium correlation
function is exactly calculated. We conclude this series of papers in
Sec.~\ref{conclusion}. Some technical details are given in Appendix
\ref{EM}-\ref{tensor1}.

Finally we list some of the notations and conventions. We choose the
unit system with the speed of light $c=1$\,. To distinguish from the
Minkowski $4$-vector we use the bold font to denote the vector in
the Euclidean space. Greek indices running from $0$ to $3$ are
further used to denote the component of the former. The Einstein
summation convention is applied to these indices. The
$4$-dimensional Minkowski space is endowed with the metric
$\eta^{\mu\nu}={\rm diag}(1,-1,-1,-1)$\,. The scalar product of two
$4$-vectors is defined as $a\cdot b \equiv \eta^{\mu\nu} a_\mu b_\nu
= a_\mu b^\mu$\,. In particular, $a\cdot a\equiv a^2$\,. In addition
to the usual mathematical symbols we use the following notations:

\begin{tabular*}{0.30\textwidth}{@{\extracolsep{\fill}}  l l   }
$\partial_\mu$\,, & covariant derivative: \, $\partial_\mu =
\partial/\partial x^\mu$ \,; \\
$d^4 z$\,, & volume element in $4$-dimensional Minkowski \\
& space: \, $d^4 z= dz^0 dz^1 dz^2 dz^3$\,;\\
$d^3 {\bf z}$\,, & volume element in $3$-dimensional Euclidean \\
& space:\, $d^3 {\bf z}= dz^1 dz^2 dz^3$\,;\\
$\delta^{(d)}(f)$\,, & $d$-dimensional
Dirac function:\\
$\theta(x)$\,, & Heaviside function\,;\\
$d\Sigma_\mu$\,, & differential form of spacelike $3$-surface:\\
& $d\Sigma_\mu = \frac{1}{3!} \epsilon_{\mu\nu\rho\lambda} \, dx^\nu
\wedge dx^\rho \wedge dx^\lambda $ with
$\epsilon_{\mu\nu\rho\lambda}$ \\
& being $ \pm 1$ when
$(\mu\nu\rho\lambda)$ is an even (odd) \\
& permutation of $(0123)$ and
being $0$ otherwise\,;\\
$x_i[\varsigma]$\,, & world line of particle $i$\,;\\
$x_i(\varsigma)$\,, & $4$-position of particle $i$ at proper time
$\varsigma$\,.
\end{tabular*}

\section{Main results of general theory}
\label{generalformalism}

This section is devoted to presenting the exact starting point that
underlies the entire analysis of the following sections. For this
purpose we first briefly review the manifestly covariant correlation
dynamics developed in Paper I, and introduce all the mathematical
objects used throughout this paper. It should be stressed that at
each step of the manipulations below the manifest covariance is
preserved.

\subsection{Correlation dynamics of $\tau$-parametrized evolution}
\label{generaltheory}

Consider a microscopic system composed of $N$ identical (classical)
particles with mass $m$\,, the dynamics of which is formulated
within the action-at-a-distance formalism \cite{Fokker29}. More
precisely, the history of the system is described by a bundle of $N$
particle world lines which solve the following relativistic motion
equations:
\begin{eqnarray}
\frac{dx^\mu_i}{d\tau_i} &=& \frac{p^\mu_i}{m}\equiv u^\mu_i\,,
\label{Newton1}\\
\frac{dp^\mu_i}{d\tau_i} &=& \sum_{j\neq i}^N F^\mu_{ij}(x_i,p_i)\,.
\label{Newton}
\end{eqnarray}
Here $x_i^\mu(\tau_i)\,, u_i^\mu(\tau_i)\,, p_i^\mu(\tau_i)$ are the
$4$-position, the $4$-velocity and the $4$-momentum vector of
particle $i$ depending on the proper time $\tau_i$\,, respectively,
and $F^\mu_{ij}$ is the force acting on particle $i$ by particle
$j$\,. Notice that $F^\mu_{ij}(x_i,p_i)$ functionally depends on the
world line of particle $j$ namely $x_j[\tau_j]$\,. For simplicity we
here consider the case where the external force is absent, and the
interacting force $F^\mu_{ij}$ is conservative, i.e.,
\begin{equation}
\frac{\partial}{\partial p_i^\mu} F^\mu_{ij}(x_i,p_i)=0 \,.
\label{forcecondition}
\end{equation}
The mass-shell constraint: $p_i^2=m^2$ is preserved by
\begin{equation}
p_i\cdot F_{ij}(x_i,p_i)=0\,.
\label{forcecondition1}
\end{equation}

Eqs.~(\ref{Newton1}) and (\ref{Newton}) suggest that in order to
formulate a statistical theory of an ensemble of such systems the
introduction of an $8N$-dimensional $\Gamma$ phase space and the
associated probability (phase) density function ${\cal D}$ is
required. (The mass-shell constraint is absorbed into the
distribution function.) Remarkably, this distribution function
differs from its Newtonian counterpart in that it is parametrized by
$N$ (rather than $1$) proper times, and functionally depends on $N$
particle world lines. For ${\cal D}$ the probability conservation
law gives $N$ manifestly covariant Liouville equations. By further
introducing an auxiliary ``gauge'' condition--to demand the $N$
proper times to change uniformly--we obtain a manifestly covariant
single-time Liouville equation of ${\cal D}(x_1,p_1,\tau_1+\tau,
\cdots
\,,x_N,p_N,\tau_N+\tau;x_{1}[\varsigma],\cdots,x_{N}[\varsigma])$\,:
\begin{eqnarray}
\left (\frac{\partial}{\partial \tau}- {\hat {\mathfrak L}}\right)
{\cal D}
=0 \,,
\label{Liouville1}
\end{eqnarray}
where the Liouvillian ${\hat {\mathfrak L}}$ is given by
\begin{eqnarray}
{\hat {\mathfrak L}}&=&{\hat {\mathfrak L}}^0 + \lambda {\hat
{\mathfrak L}}'\,, \label{interactingLiouville}\\
{\hat {\mathfrak L}}^0 &=& -\sum_{i=1}^N \,u_i^\mu
\partial_{\mu i} \,, \qquad
\lambda {\hat {\mathfrak L}}' = \sum_{i<j}\, \lambda \LI'_{ij}\,,
\nonumber\\
\lambda \LI'_{ij} &\equiv& - \left\{F^\mu_{ij}
(x_i,p_i)\frac{\partial}{\partial p_i^\mu} + F^\mu_{ji}
(x_j,p_j)\frac{\partial}{\partial p_j^\mu} \right\}\,, \nonumber
\end{eqnarray}
with ${\hat {\mathfrak L}}^0$ and $\lambda \LI'_{ij}$ the free and
the two-body interacting Liouvillian, respectively. Notice that here
the dimensionless parameter $\lambda$ characterizes the interaction
strength.

For the evolution parametrized by $\tau$ namely
Eq.~(\ref{Liouville1}) the correlation dynamics analysis may be
performed for a large class of realistic systems \cite{Tian09}.
First, we define the following distribution vector:
\begin{eqnarray}
\overrightarrow{{\frak D}} \equiv (\{{\cal D}_1\} \,, \{{\cal D}_2\}
\,, \cdots \,,\{{\cal D}_N \}\equiv {\cal D})
\,, \label{distributionvector}
\end{eqnarray}
where the reduced $s$-body distribution function ${\cal D}_s$ is
obtained by integrating out arbitrary $(N-s)$ particle phase
coordinates and because of this
for each $\{{\cal D}_s\}$ there are $N!/[(N-s!)s!]$ components. With
the help of this definition the BBGKY hierarchy is rewritten in a
compact form:
\begin{equation}
\left (\frac{\partial}{\partial \tau}- {\hat {\mathfrak
L}}\right)\overrightarrow{{\frak D}}=0 \,. \label{BBGKY1}
\end{equation}
Eqs.~(\ref{distributionvector}) and (\ref{BBGKY1}) constitute the
{\it reduced distribution function representation} of the
single-time Liouville equation (\ref{Liouville1}). To proceed
further we introduce the so-called {\it correlation pattern
representation} allowing a more delicate decomposition of the
distribution vector or more general functions. A correlation
pattern, denoted as $|\Gamma_s\rangle$ (or $\langle \Gamma_s|$),
describes the statistical correlation of given $s$-particle group.
More precisely, consider an $s$-particle group $(i_1\cdots i_s)\,,
s\leq N$\,, then the correlation pattern is generally given by
$|{\rm P_1}|{\rm P_2}|\cdots |{\rm P_j}\rangle$ (or $\langle {\rm
P_1}|{\rm P_2}|\cdots |{\rm P_j}|$), where ${\rm P}_1\,,\cdots
\,,{\rm P}_j$ is a partition of $(i_1\cdots i_s)$\,. It implies that
in the statistical sense within ${\rm P}_i\,, 1\leq i\leq j$ the
particles correlate with each other, while the particle groups ${\rm
P}_i\,, 1\leq i\leq j$ are independent. The reduced $s$-body
distribution function in this presentation now reads out as ${\cal
D}_s=\sum_{\Gamma_s}\, |\Gamma_s\rangle \langle \Gamma_s|{\cal D}_s$
which is none but the cluster expansion. It should be stressed,
however, that it differs from a traditional one
\cite{Prigogine63,Balescu} in that the distribution functions depend
on the particle world lines. This is, indeed, an important
ingredient of the Klimontovich technique in the Newtonian context
\cite{Klimontovich60} and was generalized to the special
relativity--in a manifestly covariant manner--by Hakim
\cite{Hakim67}.

A particularly important correlation pattern is the so-called {\it
vacuum state}: $|\Gamma_s\rangle \equiv |0_s\rangle$ (or $\langle
\Gamma_s| \equiv \langle 0_s|$)\,, where the given $s$ particles are
(statistically) independent. In contrast, all the other correlation
patterns are called {\it correlation states}. With this definition
the vacuum and the correlation operator, denoted as $\V$ and $\C$\,,
respectively, are defined as follows:
\begin{eqnarray}
\V\, |\Gamma_r \rangle = \delta_{0_r\Gamma_r}\, |\Gamma_r \rangle
\,, \qquad \C\, |\Gamma_r \rangle = (1-\delta_{0_r\Gamma_r})\,
|\Gamma_r \rangle \,,
\label{VCdefinition}
\end{eqnarray}
which project given functions onto the vacuum or the correlation
state.

Then, a series of rigorous theorems establish the following
important properties. First of all, the distribution vector
$\overrightarrow{{\frak D}}$ is split into the kinetic component
${\hat \Pi}_{\rm k.} \overrightarrow{{\frak D}}$ and the nonkinetic
component ${\hat \Pi}_{\rm n.k.} \overrightarrow{{\frak D}}$\,,
i.e.,
\begin{eqnarray}
\overrightarrow{{\frak D}} = {\hat \Pi}_{\rm k.}
\overrightarrow{{\frak D}} + {\hat \Pi}_{\rm n.k.}
\overrightarrow{{\frak D}} \,,
\label{decomposition}
\end{eqnarray}
and the latter is irrelevant for large global proper times. The
former is further decomposed into the vacuum and the correlation
state, i.e.,
\begin{eqnarray}
{\hat \Pi}_{\rm k.} \overrightarrow{{\frak D}} = \V {\hat \Pi}_{\rm
k.} \overrightarrow{{\frak D}} + \C {\hat \Pi}_{\rm k.}
\overrightarrow{{\frak D}} \,,
\label{decompositionVC}
\end{eqnarray}
and the correlation state is fully determined by the vacuum state.
In the thermodynamic limit $N\rightarrow +\infty$ as written in the
correlation pattern representation the evolution of $\V {\hat
\Pi}_{\rm k.} \overrightarrow{{\frak D}}$ and $\C {\hat \Pi}_{\rm
k.} \overrightarrow{{\frak D}}$ are expressed in terms of two
infinite equation hierarchies. It is remarkable that both infinite
equation hierarchies are determined merely by a reduced one-body
distribution function ${\tilde {\cal D}}(x,p;{\cal X}_{(x,p)})$
which solves the {\it exact} closed equation:
\begin{eqnarray}
&& \left\{ \frac{\partial}{\partial \tau} + u^\mu_1
\partial_{\mu 1} - \int d2 \,\lambda \LI'_{12}\,
{\tilde {\cal D}}(2;{\cal X}_2)\right\} {\tilde {\cal D}}(1;{\cal X}_1) \nonumber\\
&=& \sum_{j \geq 2}\int \! d2 \! \cdots \! \int\! dj \langle 1 |\V
(\G - \Li) \V  |1|\cdots |j\rangle \prod_{s=1}^j {\tilde {\cal
D}}(s;{\cal X}_s).\nonumber\\
\label{generalkinetic5}
\end{eqnarray}
Here we have introduced the shorthand notations: ${\tilde {\cal
D}}(i;{\cal X}_i)\equiv {\tilde {\cal D}}(x_i,p_i;{\cal
X}_{(x_i,p_i)})$ and $di\equiv d^4x_id^4p_i$\,, and the notation:
${\cal X}_{(x,p)}$ stands for some world line passing through the
phase point $(x,p)$\,. Notice that the two-body interacting
Liouvillian $\lambda\LI'_{ij}(x_i,p_i;x_j,p_j)$ is a functional of
the world lines ${\cal X}_{(x_i,p_i)}$ and ${\cal X}_{(x_j,p_j)}$\,.
The operator $\V\G\V$ determines the evolution of the vacuum state
$\V {\hat \Pi}_{\rm k.} \overrightarrow{{\frak D}}$\,, and is given
by the following functional equation:
\begin{eqnarray}
\V\G\V = \V\Li\V + \int_0^\infty ds\, \V\GG(s)\V \exp(-s\V\G\V) \,, \label{VGV}\\
\V\GG(s)\V = \int_C \frac{dz}{2\pi}\, e^{-izs} \V\E(z)\R^0(z)\C
\lambda \Li' \V \,, \qquad \nonumber
\end{eqnarray}
where the contour $C$ (in the complex plane) lies above all the
singularities of the Laplace transform of ${\cal D}$\,, and
\begin{eqnarray}
\E(z)&=&\sum_{n=0}^\infty \lambda^{n+1} \Li'\{\C\R^0(z)\Li'\}^n \,,
\label{irreducibleoperator}\\
\R^0(z) &=& \frac{1}{-iz-{\hat {\mathfrak L}}^0} \,.
 \label{UR}
\end{eqnarray}

To our best knowledge a kinetic equation similar to
Eq.~(\ref{generalkinetic5}) was first obtained by Hakim by using the
weak coupling approximation \cite{Hakim67}. A simplified equation of
Hakim has very recently been found to play important roles in
studies of the relativistic Brownian motion \cite{Hanggi09}.

\subsection{General transport equation and hierarchy
of physical correlation functions}
\label{exactequation}

The $\tau$-parametrized evolution may not be observable because
${\cal D}$ is normalized in the $8N$-dimensional phase space.
Rather, to match macroscopic observations the following physical
distribution function:
\begin{eqnarray}
{\cal N}(x_1,p_1,\cdots
\,,x_N,p_N;x_{1}[\varsigma],\cdots,x_{N}[\varsigma])
\! \equiv \! \int \!\! \prod_{i=1}^N d\tau_i {\cal D}
\label{density1}
\end{eqnarray}
is introduced which may be considered formally as the stationary
solution to Eq.~(\ref{Liouville1}). As a result, the nonkinetic
component of ${\cal N}$ identically vanishes. Moreover, in the
thermodynamic limit $N\rightarrow +\infty$ both hierarchies
(corresponding to the vacuum and the correlation state,
respectively) is determined merely by a physical one-body
distribution function $f(x,p)$\,. Here by ``physical'' the
normalization condition: $\lim_{N\rightarrow +\infty
}N^{-1}\int_{\Sigma\otimes U^4} d\Sigma_{\mu} d^4 p\, u^\mu\,
f(x,p)=1$ is implied, where $\Sigma$ is a spacelike $3$-surface, and
$U^4$ is the $4$-dimensional Minkowski momentum space.

The stationary solution to Eq.~(\ref{generalkinetic5}) results in a
general transport equation which is manifestly covariant and closed:
\begin{eqnarray}
\left\{u^\mu_1 \partial_{\mu 1}  - \int_{\Sigma_2 \otimes U^4_2 }
d\Sigma_{\mu 2} d^4 p_2u^\mu_2\lambda \LI'_{12}f(2) \right\} f(1) =
\mathbb{K} [f] \,, \nonumber\\
\label{reducedonebody}
\end{eqnarray}
where $f(i)$ is the shorthand notation of $f(x_i,p_i)$\,, and the
collision integral is given by
\begin{eqnarray}
\mathbb{K}[f] = \sum_{j \geq 2} \, \int_{\Sigma_2 \otimes U^4_2 }
d\Sigma_{\mu 2} d^4 p_2 u^\mu_2\cdots \int_{\Sigma_j \otimes U^4_j
} d\Sigma_{\mu j} d^4 p_j u^\mu_j \nonumber\\
\times \langle 1 |\V (\G - \Li) \V |1|\cdots |j\rangle \,
\prod_{i=1}^j\, f(i) \,. \qquad \qquad \label{averagekinetic10}
\end{eqnarray}
The solution to Eqs.~(\ref{reducedonebody}) and
(\ref{averagekinetic10}), in turn, uniquely determines the hierarchy
of physical correlation functions. More precisely, given an
arbitrary $j$-particle correlation pattern $\Gamma_j (1,\cdots\,,
j)\neq 0_j$\,, in the thermodynamic limit $N\rightarrow +\infty$ the
physical correlation function (denoted as $\langle \Gamma_j
|\overrightarrow{\mathfrak{N}}_\infty$) reads out as
\begin{eqnarray}
&& \langle \Gamma_j |\overrightarrow{\mathfrak{N}}_\infty
\label{correlationresult2}\\
&=& \int_0^{\infty} ds\!\!\int_0^s ds' \sum_{n=j}^\infty
\prod_{i=j+1}^n \int_{\Sigma_i \otimes U^4_i } d\Sigma_{\mu i} d^4
p_i\, u^\mu_i \, \prod_{k=1}^n f(k)
\nonumber\\
&& \times  \langle \Gamma_j
| \C {\hat {\mathfrak{U}}}^0 (s-s'){\hat {\mathfrak{E}}}(s') \V
\exp(-s\V\G\V)|1|\cdots |n\rangle \,, \nonumber
\end{eqnarray}
(The integration procedure: $\prod_{i=j+1}^n \int_{\Sigma_i \otimes
U^4_i } d\Sigma_{\mu i} d^4 p_i\, u^\mu_i$ is defined as unity for
$n=j$\,.) where
\begin{equation}
\U^0(\tau) = \int_C \frac{dz}{2\pi} \, e^{-iz\tau} \R^0(z)\,, \qquad
{\hat {\mathfrak{E}}}(\tau) = \int_C \frac{dz}{2\pi} \, e^{-iz\tau}
\E(z)\,.
 \label{URrelation}
\end{equation}

\begin{figure}
\begin{center}
\leavevmode \epsfxsize=8cm \epsfbox{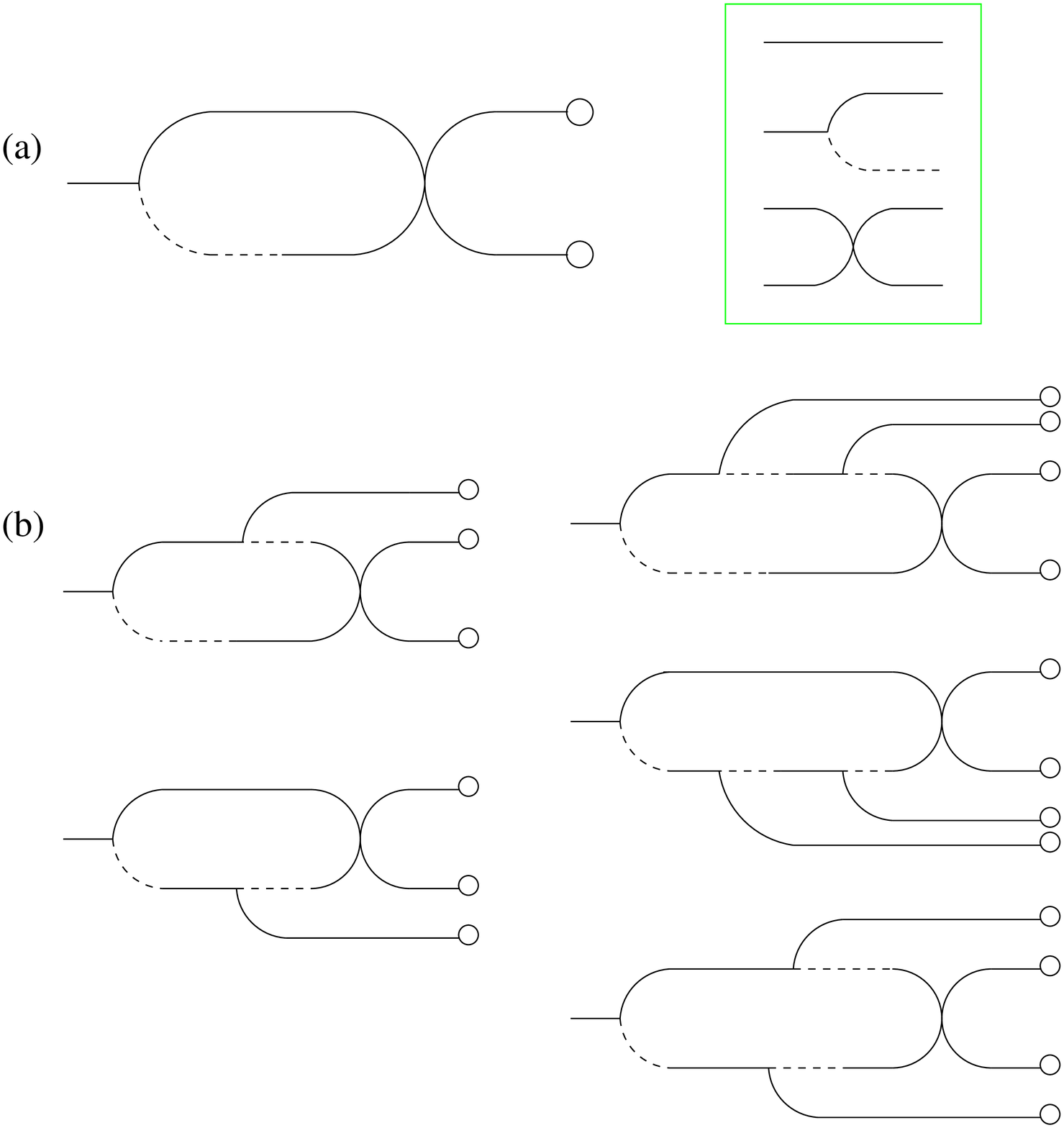}
\end{center}
\caption{Diagrams representing the perturbative expansion of the
collision integral: (a) the weak coupling approximation and (b) the
lowest order ring approximation. The circle stands for the physical
one-body distribution function. Inset: free propagator (top) and two
types of interaction vertex (middle and bottom).}
  \label{kineticfig}
\end{figure}

In practical applications a perturbative expansion with respect to
$\lambda$ may be further performed for the collision integral
$\mathbb{K}[f]$ and the physical correlation function $\langle
\Gamma_j |\overrightarrow{\mathfrak{N}}_\infty$\,. Such an expansion
is represented by the diagrams constructed out of the free
propagator and the interaction vertex (see the inset of
Fig.~\ref{kineticfig}). For the former the matrix element reads out
as
\begin{eqnarray}
\langle i'| \, e^{-\tau u^\mu_j\partial_{\mu j}}\, |i\rangle =
\delta_{ij}\delta_{ii'}\, e^{-\tau u^\mu_i\partial_{\mu i}} \,.
\label{freematrixelement}
\end{eqnarray}
For the latter there are two types: For the first type (the middle
in the inset of Fig.~\ref{kineticfig}) a particle joining the vertex
from the right is annihilated (the dashed line), and the matrix
element reads out as
\begin{eqnarray}
&& \langle i|\lambda \LI'_{i'j'}|\Gamma_2(i,j)\rangle \nonumber\\
&=&
(\delta_{ii'}\delta_{jj'}+\delta_{ij'}\delta_{ji'})\int_{\Sigma_j
\otimes U^4_j } d\Sigma_{\mu j} d^4 p_j\, u^\mu_j \,
\lambda\LI'_{ij} \label{interactionmatrixelement1}
\end{eqnarray}
with $|\Gamma_2(i,j)\rangle=|ij\rangle\, {\rm or}\, |i|j\rangle$\,.
For the second type (the bottom in the inset of
Fig.~\ref{kineticfig}) no particles (joining the vertex from the
right) are annihilated. ``Switching on'' the interaction (at the
vertex) introduces the (statistical) correlation, i.e., $\langle
\Gamma_2(i,j)| = \langle ij|$ irrespective of the ``initial''
correlation pattern (to the right of the vertex), i.e.,
$|\Gamma_2(i,j)\rangle$\,. The matrix element for this type of the
interaction vertex reads out as
\begin{eqnarray}
\langle ij|\lambda \LI'_{i'j'}|\Gamma_2(i,j)\rangle =
(\delta_{ii'}\delta_{jj'}+\delta_{ij'}\delta_{ji'})
\lambda\LI'_{ij}\,. \label{interactionmatrixelement2}
\end{eqnarray}

Eqs.~(\ref{reducedonebody})-(\ref{correlationresult2}) justify the
manifestly covariant Bogoliubov functional assumption
\cite{Prigogine63,Balescu,Bogoliubov}. They constitute the complete
set for describing various physical phenomena such as transport
processes, (macroscopic) relativistic hydrodynamics, and (physical)
correlations at equilibrium, and serve as the exact starting point
of subsequent sections. The remaining of this paper is, indeed,
devoted to the applications of
Eqs.~(\ref{reducedonebody})-(\ref{correlationresult2}) in classical
relativistic plasmas with electromagnetic interactions.

\section{Special transport equations}
\label{transport}

In this section we will consider a rarified electron plasma that is
near the local equilibrium with the density and the temperature in
the local rest frame as $\rho_0$ and $T$\,, respectively. We will
justify that various (namely Vlasov \cite{
Hakim67}, Landau \cite{Belyaev57,Klimontovich60} and Boltzmann
\cite{Cercigani02,vanLeeuwen80}) manifestly covariant kinetic
equations existing in literatures are unified within the general
transport equation (\ref{reducedonebody}). Particular attention will
be paid to the additional approximations made. In doing so we expect
to clarify the context where they are applicable.

It should be stressed that given a density $\rho_0$ the present
theory is applicable for moderate temperatures. For sufficiently
high or low temperatures quantum statistics and QED processes
dominate, and the complete treatment requires a quantum theory. Let
us now estimate such a condition. On one hand, two classical
electrons may approach each other up to a distance $\sim
e^2/(k_BT)$\,, where $k_B$ is the Boltzmann constant. In order for a
classical scattering theory to be applicable it must be larger than
the thermal deBroglie wavelength $\hbar/\sqrt{mk_B T}$\,.
Consequently, we find $k_BT \ll \alpha^2 m$ with
$\alpha=\frac{1}{137}$ the fine structure constant. On the other
hand, the plasma become degenerate when the thermal deBroglie
wavelength and the mean distance $\sim \rho_0^{-1/3}$ are
comparable. This implies that the quantum statistics can be ignored
only for sufficiently high temperatures such that $\hbar/\sqrt{mk_B
T}\ll \rho_0^{-1/3}$\,, i.e., $k_BT \gg \hbar^2 \rho_0^{2/3}/m$\,.
Thus, the temperature region for a classical theory to be applicable
is
\begin{equation}
\frac{\hbar^2 \rho_0^{2/3}}{m} \ll k_B T \ll \alpha^2 m \,.
\label{Tregion}
\end{equation}
Notice that the inequality above imposes a restriction on the
density, i.e., $\rho_0\ll (e^2m/\hbar^2)^3$\,. This implies that the
fermi energy namely $\hbar^2 \rho_0^{2/3}/m$ is the lowest energy
scale and, in particular, is much smaller than the (classical)
electromagnetic energy $e^2 \rho_0^{1/3}$\,. The inequality
(\ref{Tregion}) is the rigorous condition for the present classical
theory to be applicable.

\subsection{General scheme}
\label{scheme}

Let us first prove an exact relation between the collision integral
$\mathbb{K}[f]$ and the physical two-body correlation function. For
the latter setting $\langle \Gamma_j|$ to be $\langle 12|$ for the
hierarchy (\ref{correlationresult2}) we find
\begin{eqnarray}
&& \langle 12 |\overrightarrow{\mathfrak{N}}_\infty
\label{correlation3}\\
&=& \int_0^{\infty} ds\!\!\int_0^s ds' \sum_{n=j}^\infty
\prod_{i=j+1}^n \int_{\Sigma_i \otimes U^4_i } d\Sigma_{\mu i} d^4
p_i\, u^\mu_i \, \prod_{k=1}^n f(k)
\nonumber\\
&& \times  \langle 12
| \C {\hat {\mathfrak{U}}}^0 (s-s'){\hat {\mathfrak{E}}}(s') \V
\exp(-s\V\G\V)|1|\cdots |n\rangle \,.
\nonumber
\end{eqnarray}
Notice that
\begin{eqnarray}
\V(\G-\Li)\V = \qquad \qquad \qquad \qquad \label{VGV}\\
\int_0^\infty ds\int_0^s ds'\V\lambda \Li' \C {\hat
{\mathfrak{U}}}^0 (s-s')\E(s') \V \exp(-s\V\G\V) \,. \nonumber
\end{eqnarray}
Inserting it into Eq.~(\ref{averagekinetic10}) gives
\begin{eqnarray}
\mathbb{K}[f] =\int_{\Sigma_2 \otimes U^4_2 } d\Sigma_{\mu 2} d^4
p_2\, u^\mu_2 \langle 1| \lambda {\hat {\mathfrak{L}}}'|12 \rangle \langle 12
|\overrightarrow{\mathfrak{N}}_\infty\,.
\label{kineticria1}
\end{eqnarray}
This exact relation indicates that the partition of full
distribution functions in the correlation pattern representation
preserves the cluster expansion, which is guaranteed by the Clavin
theorem in the Newtonian physics \cite{Clavin72}.

We remark that from Eq.~(\ref{kineticria1}) the collision integral
is locally well defined provided that either interactions or
(statistical) correlations are short-ranged in the spacelike
$3$-surface passing through $x_1$ and $x_2$\,. A relativistic plasma
with electromagnetic interactions, indeed, belongs to the latter
case. There, although the static transverse electromagnetic field is
long-ranged, the Debye screening of the longitudinal electromagnetic
field, as we will show in Sec.~\ref{equilibrium}, renders the
correlation function short-ranged with the correlation radius
\begin{eqnarray}
\lambda_{\rm D}= \sqrt{\frac{k_BT}{4\pi e^2\rho_0}} \,.
\label{ringcorrelation2}
\end{eqnarray}
As a result, in Eq.~(\ref{kineticria1}) the integral over
$d\Sigma_{\mu 2}$ is dominated by the region around $x_1$ of size
$\lambda_{\rm D}$\,. Notice that the correlation radius
Eq.~(\ref{ringcorrelation2}) makes sense only if it is much larger
than the mean distance between two nearest electrons, which is order
of $\rho_0^{-1/3}$\,. This leads to a sufficiently small plasma
parameter, i.e., $e^2\rho_0^{1/3}/k_B T\ll 1$\,. Combining with the
inequality (\ref{Tregion}) we obtain
\begin{equation}
\frac{\hbar^2 \rho_0^{2/3}}{m} \ll e^2\rho_0^{1/3} \ll k_B T \ll
\alpha^2 m \,, \label{Tregion1}
\end{equation}
which is the exact condition for the subsequent analysis to be
applicable.

Then, the low density limit introduces a substantial simplification
of the collision integral ${\mathbb{K}}[f]$\,. Indeed, because the
physical one-body distribution is proportional to the density in
Eq.~(\ref{reducedonebody}) the collision integral may be considered
formally as a density expansion which, term by term, corresponds to
the two-, three-body scattering and so on. In the low density limit
all the higher order terms in this density expansion may be ignored.
As a result,
\begin{eqnarray}
\mathbb{K}[f] &=& \int_{\Sigma_2 \otimes U^4_2 } d\Sigma_{\mu 2} d^4
p_2 u^\mu_2 \nonumber\\
&& \qquad \quad \times \langle 1| \V(\G-\Li)\V |1|2\rangle \,
f(1)f(2)\,. \label{kineticria3}
\end{eqnarray}
Furthermore, with the help of appropriate iteration for
Eq.~(\ref{VGV}) we explicitly write down the operator $\V\G\V$ as
\cite{Balescu}
\begin{eqnarray}
\V (\G - \Li) \V = \sum_{n=1}^\infty \, \V\G_{[n]}\V \,.
\label{VGV2}
\end{eqnarray}
Here
\begin{widetext}
\begin{eqnarray}
\G_{[n]} &=& \int_0^\infty\!\! ds_2\int_0^\infty\!\! ds_4\cdots
\int_0^\infty\!\! ds_{2n} \int_0^{s_1-s_2}\!\!
ds_3\int_0^{s_3-s_4}\!\! ds_5\cdots \int_0^{s_{2n-3}-s_{2n-2}}\!\!
ds_{2n-1}\, \V\GG(s_2)\V \U(s_1-s_2-s_3)\V \nonumber\\
&& \qquad \qquad \times \GG(s_4)\V \U(s_3-s_4-s_5)\V \cdots \GG(s_{2n})\V
\U(s_{2n-1}-s_{2n}-s_{2n+1})\V\,,
\label{VGV1}
\end{eqnarray}
\end{widetext}
where $s_1=s_{2n+1}=0$\,, and the operator $\V\U(s)\V=\V
\exp\{s\V\Li\V\}$\,. Notice that in order for the matrix element
$\langle \Gamma_r |\V\GG(s)\V| \Gamma'_{r'} \rangle$ not to vanish
the condition: $\Gamma_r = 0_r\,, r\geq 2$ and $\Gamma_{r'} =
0_{r'}\,, r'> 2$ must be met. For $\G_{[n]}$ with $n\geq 2$ there
are more than one particles annihilated, which leads to higher order
density corrections as the interaction strength $\lambda$ is
compensated by a density factor associated with the annihilated
particle. For this reason in the expansion of Eq.~(\ref{VGV2}) only
the leading term ($n=1$) is kept. Furthermore, because in the
$\lambda$-expansion of $\V\U\V$ the higher order terms are
associated with the particle annihilation resulting in higher-order
density corrections, the replacement: $\U(-s) \rightarrow \U^0(-s)$
may be made. Consequently, we obtain
\begin{eqnarray}
\mathbb{K}[f] &=& \int_{\Sigma_2 \otimes U^4_2 } d\Sigma_{\mu 2} d^4
p_2\, u^\mu_2 \label{kineticria4}\\
&& \times \int_0^\infty\!\! ds \langle 1 |\V \GG(s)\V \U^0(-s) \V
|1|2\rangle\, f(1)f(2)
\nonumber
\end{eqnarray}
as the (formal) leading order density expansion of the collision
integral (\ref{kineticria3}) concerned.

The simplified collision integral (\ref{kineticria4}) can be
expressed in terms of the $\lambda$-expansion. It is important that
for this expansion the interaction strength $\lambda$ is {\it not}
compensated by the density factor because no particles are
annihilated. In the remaining part we show that keeping such an
expansion up to the $\lambda/\lambda^2$ term results in the
manifestly covariant Vlasov/Landau equation, while keeping the
entire expansion results in the manifestly covariant Boltzmann
equation.

Finally let us present a summary of the approximations to be used in
the subsequent analysis that implement the scheme outlined above.
One is the so-called {\it relativistic impulse approximation}
\cite{Israel84,Kandrup84}. There, the phase trajectory ${\cal
X}_{(x,p)}$ in Eqs.~(\ref{reducedonebody}) and
(\ref{averagekinetic10}) is given by
\begin{equation}
{\cal X}_{(x,p)} \equiv x[s]=x^\mu - u^\mu s \,, \label{impulse}
\end{equation}
where we choose the proper time origin to be the moment at which the
world line passes through $x$ with the given $4$-velocity $u^\mu$\,.
The other is the traditional {\it hydrodynamic approximation}
assuming that the physical distribution function $f(x,p)$ varies
over a spatial (temporal) scale much larger than $\lambda_{\rm D}$
($\omega_p^{-1}=\sqrt{m/4\pi e^2 \rho_0}$)\,.
Eq.~(\ref{kineticria4}), in combination with these two
approximations, is the starting point of subsequent analysis.

\subsection{Mean field approximation: Relativistic Vlasov equation}
\label{Vlasov}

\begin{figure}
\begin{center}
\leavevmode \epsfxsize=8cm \epsfbox{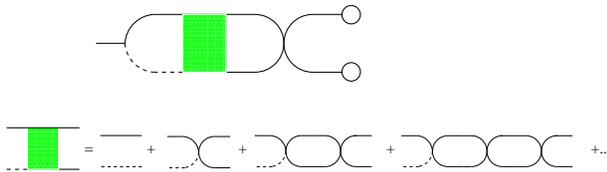}
\end{center}
\caption{Diagrams leading to the Boltzmann collision integral.}
  \label{kineticfig1}
\end{figure}

Let us start from the simplest case namely to keep the
$\lambda$-expansion up to the first order. That is, we neglect the
collision integral. Consequently, we obtain
\begin{eqnarray}
\left\{ u^\mu_1 \partial_{\mu 1} - \lambda \int_{\Sigma_2 \otimes
U^4_2 } d\Sigma_{\mu 2} d^4 p_2\, u^\mu_2 \LI'_{12} \, f(2) \right\}
f(1) = 0. \label{firstorder}
\end{eqnarray}
According to the second term of Eq.~(\ref{firstorder}), the physical
one-body distribution function is driven by the mean field formed by
all the other particles. For this reason, to keep the interaction
expansion up to the leading order is called {\it mean field
approximation}. To fully determine the mean field we use the
relativistic impulse approximation Eq.~(\ref{impulse}). As a result,
\begin{widetext}
\begin{eqnarray}
&& \lambda \LI'_{12} (x_1,p_1;x_2,p_2) \approx  \frac{i 8\pi^2
e^2}{m} \, \int \frac{d^4
k}{(2\pi)^4}\, e^{ ik \cdot (x_1-x_2)} \, {\hat {\cal G}}_{12}(k) \,, \nonumber\\
{\hat {\cal G}}_{12}(k) & \equiv & \frac{1}{k^2}\, \left\{\delta
(k\cdot p_2) [k^\mu (p_1\cdot p_2)-p_2^\mu (k\cdot p_1)]
\frac{\partial}{\partial p^\mu_1} - \delta (k\cdot p_1) [k^\mu
(p_1\cdot p_2)-p_1^\mu (k\cdot p_2)] \frac{\partial}{\partial
p^\mu_2}\right\} \,. \label{forceEMresult}
\end{eqnarray}
The derivation is given in Appendix~\ref{impulseLiouville}.

\subsection{Weak coupling approximation: Relativistic Landau equation}
\label{BBK}

In this part we will consider the so-called {\it weak coupling
approximation} to Eq.~(\ref{kineticria4}). That is, the
$\lambda$-expansion is kept up to the second order
[Fig.~\ref{kineticfig} (a)]. The collision integral thereby obtained
is denoted as $\mathbb{K}_1[f]$\,.

\subsubsection{Collision integral}
\label{BBK1}

Under the weak coupling approximation Eq.~(\ref{kineticria4}) is
simplified as
\begin{eqnarray}
\mathbb{K}_1[f] = \lambda^2 \int_0^\infty \!\! ds \!\!
\int_{\Sigma_2 \otimes U^4_2 } d\Sigma_{\mu 2} d^4 p_2\, u^\mu_2
{\hat {\cal L}}'_{12} {\hat {\bf {\rm U}}}^0_{12}(s) {\hat {\cal
L}}'_{12} {\hat {\bf {\rm U}}}^0_{12}(-s) f(x_2,p_2) f(x_1,p_1)\,,
\label{collision}
\end{eqnarray}
where the propagator ${\hat {\rm U}}^0_{ij} (s)$ is defined as
\begin{eqnarray}
{\hat {\rm U}}^0_{ij} (s) = \exp [-s(u_i^\mu \partial_{\mu
i}+u_j^\mu
\partial_{\mu j})] \,.
\label{Ufree}
\end{eqnarray}

As discussed in Sec.~\ref{scheme} because of the short-ranged
two-body correlation the integration over $d\Sigma_{\mu 2}$ is
dominated by a region of size $\lambda_{\rm D}$\,. Applying the
hydrodynamic approximation gives
\begin{equation}
\int_{\Sigma_2 \otimes U^4_2 } d\Sigma_{\mu 2} d^4 p_2\, u^\mu_2\,
(\cdots) f(x_1,p_1) f(x_2,p_2) \approx \int_{\Sigma_2 \otimes U^4_2
} d\Sigma_{\mu 2} d^4 p_2\, u^\mu_2\, (\cdots) f(x_1,p_1) f(x_1,p_2)
\label{localization}
\end{equation}
for the collision integral $\mathbb{K}_1[f]$\,. That is, the
collision integral is local in spacetime:
\begin{eqnarray}
\mathbb{K}_1[f] = \lambda^2 \int_0^\infty \!\! ds \!\!
\int_{\Sigma_2 \otimes U^4_2 } d\Sigma_{\mu 2} d^4 p_2\, u^\mu_2\,
{\hat {\cal L}}'_{12} {\hat {\bf {\rm U}}}^0_{12}(s) {\hat {\cal
L}}'_{12} {\hat {\bf {\rm U}}}^0_{12}(-s) f(x_1,p_2) f(x_1,p_1) \,.
\label{collision1}
\end{eqnarray}
With the relativistic impulse approximation further used we obtain
\begin{eqnarray}
\mathbb{K}_1[f] &=& \frac{\partial}{\partial p_1^\mu} \, \int d^4
p_2 \,
\epsilon^{\mu\nu} \,
\left(\frac{\partial}{\partial p_1^\nu}-\frac{\partial}{\partial
p_2^\nu}\right) \, f(x_1,p_2) f(x_1,p_1)\,,
\label{BBKcollisionresult} \\
\epsilon^{\mu\nu} &=& 2e^4\,(u_1\cdot u_2)^2 \int d^4 k\,
\delta(k\cdot u_1)\, \delta(k\cdot u_2)\, \frac{k^\mu k^\nu}{(k\cdot
k)^2} \label{tensor2}
\end{eqnarray}
after straightforward but tedious calculations, which are detailed
in Appendix~\ref{Landau}. Eq.~(\ref{BBKcollisionresult}) is the
relativistic Landau collision integral \cite{Belyaev57}. It was
first justified by Klimontovich at the full microscopic level albeit
in a nonmanifestly covariant manner
\cite{Klimontovich60,Klimontovich67}. Notice that although the
relativistic Landau collision integral is divergent, it formally
admits the J{\"u}ttner distribution as the unique (local)
equilibrium distribution (see Sec.~\ref{Juettner1} for detailed
analysis), and gives the relaxation time (up to a numerical factor)
as $\sqrt {m (k_BT)^3}/ (e^4\rho_0) \gg \omega_p^{-1}$\,.

\subsubsection{Logarithmic divergence of the collision integral}
\label{BBK2}

However, the collision integral $\mathbb{K}_1[f]$ suffers from both
the infrared and the ultraviolet divergence. Indeed, with the
integral over wave vector carried out Eq.~(\ref{tensor2}) gives (see
Appendix~\ref{tensor1} for details)
\begin{eqnarray}
\epsilon^{\mu\nu} &=& -2\pi e^4 \,\int \frac{dk_\perp}{k_\perp}\,
(u_1\cdot u_2)^2 [(u_1\cdot
u_2)^2-1]^{-3/2} \nonumber\\
&& \qquad \qquad \qquad \times \left\{[(u_1\cdot u_2)^2-1]
g^{\mu\nu}+(u_1^\mu u_1^\nu+u_2^\mu u_2^\nu)-(u_1\cdot u_2)(u_1^\mu
u_2^\nu+u_2^\mu u_1^\nu)\right\} \,.
\label{tensor}
\end{eqnarray}
\end{widetext}
A similar logarithmic divergence of the collision integral was first
noticed by Landau \cite{Landau37} in the context of nonrelativistic
plasmas (the so-called {\it Coulomb logarithm}). Following the
prescription of Landau, from the practical viewpoint, in order to
describe nonequilibrium processes near (local) J{\"u}ttner
equilibrium it suffices to substitute appropriate ultraviolet
(infrared) cutoff $k_{\rm max}$ ($k_{\rm min}$) into the collision
integral since the divergence is logarithmic. To further estimate
these cutoffs in the relativistic context we notice that
$k^{-1}_{\rm max}$ is the minimal distance as two classical
electrons approach each other. At such a distance the kinetic energy
and the interaction becomes comparable, i.e.,
\begin{equation}
k_{\rm max} \sim \frac{k_B T}{e^2} \,.
    \label{upper}
\end{equation}
For moderate temperatures $k_B T\ll \alpha^2 m$ to heal the
ultraviolet divergence is well within the reach of the present
theory. Physically, the weak coupling approximation namely the
leading order $\lambda$-expansion of Eq.~(\ref{kineticria4})
accounts for the small angle scattering, but fails in describing
large angle scattering. To implement this one needs to sum up the
entire $\lambda$-expansion of Eq.~(\ref{kineticria4}) or the
diagrams shown in Fig.~\ref{kineticfig1}. (Let us keep in mind that
the collision integral thereby obtained {\it formally} is the first
order density expansion of $\mathbb{K}[f]$\,.) This is, indeed, the
main issue of Sec.~\ref{Boltzmann}. (For higher temperatures the QED
scattering processes become dominant, and the complete treatment
must be built on a quantum theory.)

The infrared divergence reflects the long-ranged nature of
electromagnetic interactions. The present case crucially differs
from the nonrelativistic case in that the divergence exists also for
the transverse electromagnetic interaction [see Eq.~(\ref{Q}) below]
which is, however, only dynamically screened \cite{Litim02,Landau8}.
As a result, the infrared divergence has to be healed by the
short-ranged correlation, which is to be detailed in
Sec.~\ref{correlation4} and was first noticed by Klimontovich in the
formulation of a nonmanifestly covariant theory
\cite{Klimontovich67}. The infrared cutoff is therefore set by
inverse correlation radius, i.e.,
\begin{equation}
k_{\rm min} \sim \lambda_{\rm D}^{-1} \,.
    \label{lower}
\end{equation}

The infrared divergence of the Landau collision integral roots in
that the weak coupling approximation fails to capture the collective
dielectric effects which leads to the short-ranged two-body
correlation. This may be most readily seen by writing the
relativistic Landau equation in the observer's frame. For this
purpose, let us introduce the distribution function $f_0(x,p)$
defined as
\begin{equation}
f(x,p)\equiv f_0(x,p) 2m \theta(p^0)\delta (p^2-m^2) \,. \label{f}
\end{equation}
To simplify discussions we consider the spatially homogeneous case.
[The resulting $f_0$ is denoted as $f_0({\bf p},t)$\,.] Then, the
kinetic equation (\ref{reducedonebody}) with the collision integral
$\mathbb{K}_1[f]$ can be exactly rewritten as \cite{Klimontovich67}
\begin{equation}
\frac{\partial}{\partial t}f_0({\bf p},t)=\frac{\partial}{\partial
{\bf p}}\cdot \int d{\bf p}' {\bf Q}\cdot \left(\frac{\partial
}{\partial {\bf p}}-\frac{\partial }{\partial {\bf
p}'}\right)f_0({\bf p},t)f_0({\bf p}',t)\,.
\label{LandauLaboratorytframe}
\end{equation}
Here the tensor (in the $3$-dimensional Euclidean space) ${\bf Q}$
reads out as (${\bf v}$ and ${\bf v}'$ below are the velocity.)
\begin{eqnarray}
{\bf Q} = 2e^4 \int d\omega \!\!\int d^3{\bf k}\, \delta(\omega-{\bf
k}\cdot {\bf v})\delta(\omega-{\bf k}\cdot {\bf
v}') \nonumber\\
\quad \times \frac{{\bf k}{\bf k}}{|{\bf
k}|^4}\left\{\frac{1}{|\varepsilon^\parallel(\omega,{\bf k})|^2}
+\frac{[({\bf k}\times {\bf v})\cdot ({\bf k}\times {\bf v}')]^2}
{|\omega^2 \varepsilon^\perp(\omega,{\bf k})-{\bf k}^2|^2 } \right\}
\,,
\label{Q}
\end{eqnarray}
where $\varepsilon^\parallel(\omega,{\bf
k})=\varepsilon^\perp(\omega,{\bf k})=1$\,. The second term in the
curl bracket arises from the interaction mediated by the transverse
electromagnetic field. It is negligibly small in the nonrelativistic
limit, i.e., $|{\bf v}|\,, |{\bf v}'|\ll 1$\,, while the first term
arises from the interaction mediated by the longitudinal
electromagnetic field and possesses the general structure of the
Landau-Balescu-Lenard equation \cite{Landau37,Balescu60}. In
comparing with the nonmanifestly kinetic equation obtained by
Klimontivich \cite{Klimontovich67} Eq.~(\ref{Q}) suggests that the
longitudinal (transverse) permittivity $\varepsilon^\parallel$
($\varepsilon^\perp$) is unity and, thus, the collective dielectric
response is completely ignored. Nevertheless this is an artifact of
the weak coupling approximation. In fact, the nonmanifestly
covariant theory \cite{Klimontovich67} has shown
\begin{eqnarray}
\varepsilon^\parallel(\omega\rightarrow 0,{\bf k})-1 &\propto&
\frac{1}{(|{\bf k}|\lambda_{\rm D})^2} \,, \nonumber\\
\varepsilon^\perp(\omega\rightarrow 0,{\bf k})-1 &\propto&
\frac{ie^2}{\omega |{\bf k}|} \,, \label{permittivity}
\end{eqnarray}
which suggests that the fluctuations of the longitudinal
(transverse) electromagnetic field acquire a static (dynamical) mass
$\propto e^2$ (namely the interaction strength). By inserting
Eq.~(\ref{permittivity}) into Eq.~(\ref{Q}) it is clear that for a
collision integral to account for the collective dielectric effects
to go beyond the weak coupling approximation is inevitable. More
precisely, (in the Newtonian physics) the collective dielectric
effect is well known to result in a short-ranged two-body
correlation, and to be responsible for by more complicated
correlation associated with higher order terms in the density
expansion of the collision integral \cite{Balescu60}.

In the present context to treat the infrared divergence accurately
we need to sum up all the so-called {\it ring diagrams} (to be
defined in Sec.~\ref{correlation4}). In Fig.~\ref{kineticfig} (b) we
present the lowest order ring diagrams. In doing so we expect to
obtain a manifestly covariant generalization of the Balescu-Lenard
equation, where the infrared divergence is healed by collective
dielectric effects. However, to carry out this program is far beyond
the scope of the present work. We here limit ourselves to confirm
this important observation in the case of global equilibrium, which
is detailed in Sec.~\ref{correlation4}.

\subsection{Two-body scattering approximation: Relativistic Boltzmann equation}
\label{Boltzmann}

In this part we wish to justify the relativistic Boltzmann equation.
The derivation below may be generalized to low density systems with
short-ranged interactions, where the two-body scattering dominates.

\subsubsection{Formal collision integral}
\label{HakimBoltzmann}

We first introduce the following propagator:
\begin{equation}
\Y_{ijz}= \sum_{n=0}^\infty \lambda^n \langle ij| \C\, \R^0(z) [\Li'
\C\, \R^0(z)]^n |ij\rangle \,. \label{U}
\end{equation}
Upon passing to the time representation: $\Y_{ijz}\rightarrow
\Y_{ij}(s)$\,, we find
\begin{equation}
\frac{\partial}{\partial s}\Y_{ij}(s) = \left({\hat {\cal L}}^0_i +
{\hat {\cal L}}^0_j + \lambda {\hat {\cal L}}'_{ij}\right)
\Y_{ij}(s) \,, \Y_{ij}(0)=1 \,. \label{Y}
\end{equation}
Then, for Eq.~(\ref{kineticria4}) we keep the entire
$\lambda$-expansion (Fig.~\ref{kineticfig1}) which is called the
{\it two-body scattering approximation}. The collision integral
thereby obtained is denoted as $\mathbb{K}_2[f]$ which reads out as
\begin{eqnarray}
\mathbb{K}_2[f]&=& \lambda^2 \int_0^\infty \!\! ds \!\! \int_{\Sigma_2
\otimes U^4_2 } d\Sigma_{\mu 2} d^4 p_2\, u^\mu_2 \label{Boltzmann3}\\
&& \times {\hat {\cal L}}'_{12} \Y_{12}(s) {\hat {\cal L}}'_{12}
{\hat {\bf {\rm U}}}^0_{12}(-s) f(x_2,p_2) f(x_1,p_1)\,.
\nonumber
\end{eqnarray}
It differs from Eq.~(\ref{collision1}) in that the propagator ${\hat
{\bf {\rm U}}}^0_{12}(s)$ sandwiched by ${\hat {\cal L}}'_{12}$ is
renormalized into $\Y_{12}(s)$ (Fig.~\ref{kineticfig1}, lower
panel).

The derivation is exact so far. In the leading order hydrodynamic
expansion the spacetime inhomogeneity of $f(x,p)$ and two-body
scattering are decoupled. That is, we may also apply
Eq.~(\ref{localization}) to the collision integral $\mathbb{K}_2[f]$
and, furthermore, approximate the operator ${\hat {\bf {\rm
U}}}^0_{12}(-s) $ in $\mathbb{K}_2[f]$ by unity. As a result, with
$s$ integrated out Eq.~(\ref{Boltzmann3}) is simplified as
\begin{eqnarray}
\mathbb{K}_2[f]&=& \lambda^2 \int_{\Sigma_2 \otimes U^4_2 }
d\Sigma_{\mu 2} d^4 p_2\, u^\mu_2 \label{Boltzmann4}\\
&& \qquad \qquad \times {\hat {\cal L}}'_{12}\, \Z \,
{\hat {\cal L}}'_{12}\, f(x_1,p_2) f(x_1,p_1)\,, \nonumber
\end{eqnarray}
where $\Z=\int_0^\infty ds\, \Y_{12}(s)$\,. In order to find $\Z$ we
use Eq.~(\ref{Y}) to set up the following equation:
\begin{equation}
\Z={\hat G}+{\hat G} \, \lambda \hL\,\Z\,, \qquad {\hat
G}=\int_0^\infty ds {\hat {\bf {\rm U}}}^0_{12}(s) \,. \label{Z}
\end{equation}

\subsubsection{Two-body scattering and collision integral}
\label{scattering}

As before we employ the relativistic impulse approximation. There we
insert Eq.~(\ref{impulse}) into the two-body interacting Liouvillian
$\lambda \LI'_{12}$\,. To proceed further we notice that to describe
the free motion of particles $1$ and $2$\,, with the $4$-momentum
vector $p_1$ and $p_2$\,, respectively, we may work in the
center-of-momentum (CM) frame (Correspondingly we use the prime to
denote vectors in this frame.) where ${\bf p}'_1+{\bf p}'_2 =0$\,.
Because the correlation radius is short-ranged they do not interact
with each other until at a distance of $\sim \lambda_{\rm D}$\,,
when they start to repel each other. Eventually as their deviation
reaches order of $\lambda_{\rm D}$ they undergo free flight again.
Then, we assume that in the CM frame the interaction between two
particles is switched on simultaneously. This, indeed, is perfectly
legitimate because the hydrodynamic approximation washes out effects
arising from the small (coordinate) time mismatch $\sim \lambda_{\rm
D}$ of two particles. (Such an assumption may be released if we
simplify short-ranged interactions as point-like collisions. In the
latter case it can be shown that the derivations below are exact.)
As such we find that, in the CM frame,
\begin{eqnarray}
{\bf p}'_1(t')+{\bf p}'_2(t') =0 \label{CM}
\end{eqnarray}
for all $t'$\,, moreover, the two-body interacting Liouvillian does
not depend on the coordinate time $t'$ any more:
\begin{equation}
\lambda \LI'_{ij} (x'_i,p'_i;x'_j,p'_j) \equiv \lambda \LI'_{ij}
({\bf x}'_i-{\bf x}'_j,p'_i,p'_j) \,. \label{impulseLiouville1}
\end{equation}
That is, in such frame the two-body
dynamics may be reduced into the dynamics of single particle subject to some
external time-independent potential.
The detailed analysis is presented in Appendix~\ref{thirdlaw}.

Let us introduce the following notations: $u'= p'_1/m
\equiv(u'^0,{\bf u}'),\, r'\equiv (t',x',y',z')\equiv (t_1'=t_2',
({\bf x}_1'-{\bf
x}_2')/2)
$ assuming that the particle $1$ approaches to (departures from)
particle $2$ from $z'=-\infty$ (at $z'=+\infty$). Since $\lambda \LI'_{12}$
is time-independent in the CM frame, from Eq.~(\ref{Boltzmann4}) we see that in the leading
order hydrodynamic expansion the derivative with respect to $t'$ involved in the definition
of ${\hat G}$\,, namely Eq.~(\ref{Z}) may be dropped out. As a result, the
matrix element of ${\hat G}$\,, denoted as $G({\bf r}',p';\tilde{{\bf r}}',{\tilde
p}')$ satisfies
\begin{equation}
2 {\bf u}'\cdot \frac{\partial}{\partial {\bf r}'}\, G({\bf r}',p';{\tilde
{\bf r}}',{\tilde p}')=\delta^{(3)}({\bf r}'-{\tilde {\bf r}}')\delta^{(4)}(p'-{\tilde
p}')\,. \label{GCM}
\end{equation}
Using this equation it can be checked that the matrix element of $\Z$\,,
denoted as $Z({\bf r}',p';\tilde{{\bf r}}',{\tilde
p}')$ solves
\begin{eqnarray}
\left\{ 2 {\bf u}'\cdot \frac{\partial}{\partial {\bf r}'} -\lambda \LI'_{12}
\right\} Z({\bf r}',p';\tilde{{\bf r}}',{\tilde
p}')\nonumber\\
=\delta^{(3)}({\bf r}'-{\tilde
{\bf r}}')\delta^{(4)}(p'-{\tilde p}')\,. \quad
\label{ZCM}
\end{eqnarray}
Analogous to quantum mechanics Eq.~(\ref{ZCM}) describe the
scattering of an incident ``wave function'' $\psi_{\rm in} (p')$
under the ``potential'' which now reads out as $-\lambda
\LI'_{12}$\,. And the out-going ``wave function'' $\psi_{\rm out}
({\bf r}',p')$ is
\begin{eqnarray}
\psi_{\rm out} ({\bf r}',p') = \psi_{\rm in} (p') \qquad \qquad
\qquad \qquad \qquad \qquad \qquad \label{scatteredwave}\\
+ \lambda
\int d^3{\tilde {\bf r}}' \int_{U^4} d^4 {\tilde p}'
{\tilde u}'^0\, Z({\bf r}',p';{\tilde {\bf r}}',{\tilde p}') \LI'_{12}
\psi_{\rm in} ({\tilde p}') \,. \nonumber
\end{eqnarray}
Using Eq.~(\ref{ZCM}) we find
\begin{equation}
\left\{ 2 {\bf u}'\cdot \frac{\partial}{\partial {\bf r}'} -\lambda
\LI'_{12} \right\}\psi_{\rm out} ({\bf r}',p')=0\,,
\label{stationaryscatter}
\end{equation}
In deriving this equation we keep in mind that the two-body
scattering is a local event with a characteristic scale
$\lambda_{\rm D}$\,. Over such a scale incident wave functions are
strongly scattered (by the potential $-\lambda \LI'_{12}$) and, as
such, the out-going wave function generally acquires a strong
dependence on ${\bf r}'$\,. In contrast, as the two-body scattering
(in the CM frame) concerned the incident wave function $\psi_{\rm
in}$ in Eq.~(\ref{scatteredwave}) may be regarded as a spacetime
independent object. Nevertheless we notice that this is not true in
the observer's frame where, instead, the incident wave function
reads out as
\begin{equation}
\psi_{\rm in} (p') = f(x_1,p_1)f(x_1,p_2) \label{incoming}
\end{equation}
varying over a macroscopic scale $\gtrsim \lambda_{\rm D}$\,. In
Eq.~(\ref{incoming}) $p'$ is uniquely determined by $p_{1,2}$\,.

Substituting Eq.~(\ref{scatteredwave}) into Eq.~(\ref{Boltzmann4})
we find that, in terms of the CM frame
coordinates, the collision integral is
\begin{eqnarray}
\mathbb{K}_2[f] &=& \lambda \int d^3{\bf r}'\int_{U^4} d^4 p' u'^0 \,
\LI'_{12} \left\{\psi_{\rm out} ({\bf r}',p') - \psi_{\rm
in} (p')\right\} \nonumber\\
&=& \lambda \int d^3{\bf r}'\int_{U^4} d^4 p' u'^0 \, \LI'_{12} \psi_{\rm out} ({\bf r}',p') \nonumber\\
&=& \int_{U^4} d^4 p'\int d^3{\bf r}' \, 2u'^0{\bf u}'\cdot
\frac{\partial}{\partial {\bf r}'} \psi_{\rm out} ({\bf r}',p') \,,
\label{I2result}
\end{eqnarray}
where in deriving the second line we use the fact that
$\int d^3 {\bf r}' \LI'_{12}=0$\,, and in deriving
the third line we use Eq.~(\ref{stationaryscatter}).

\subsubsection{Boltzmann collision integral}
\label{Boltzmann5}

For Eq.~(\ref{I2result}) let us integrate out ${\bf r}'$ first. For
this purpose we fix the $z'$-axis to be in the direction of ${\bf
u}'$\,. Then,
\begin{widetext}
\begin{eqnarray}
&& \int d^3{\bf r}'\, 2u'^0{\bf u}'\cdot \frac{\partial}{\partial
{\bf r}'} \psi_{\rm out} ({\bf r}',p') = \int d^3{\bf r}'\,
2u'^0|{\bf u}'|\frac{\partial}{\partial z'} \psi_{\rm out} ({\bf
r}',p')
\label{I2result1}\\
&&\qquad \qquad \qquad =\int dx'dy' \{u'^0 |{\bf u}'|-u'^0 (-|{\bf
u}'|)\}\, \left\{\psi_{\rm out} ({\bf r}',p')|_{z'\rightarrow
+\infty}-\psi_{\rm out} ({\bf r}',p')|_{z'\rightarrow
-\infty}\right\} \,. \nonumber
\end{eqnarray}
\end{widetext}
Remarkably, Eq.~(\ref{I2result1}) depends only on the wave functions
at $z'=\pm \infty$ and is irrespective of the details of the
two-body scattering. Noticing that the interaction vanishes at
$z'\rightarrow \pm \infty$ we thus find $\psi_{\rm out} ({\bf
r}',p')|_{z'\rightarrow -\infty} = \psi_{\rm in} (p') $\,. In the CM
frame applying the Liouville theorem further gives
\begin{eqnarray}
\psi_{\rm out} ({\bf r}',p')|_{z'\rightarrow +\infty} &=& \psi_{\rm
in} ({\tilde p}'(\Omega'))\nonumber\\
&=& f(x_1,{\tilde p}_1)f(x_1,{\tilde
p}_2)\,.
\label{outgoingwave}
\end{eqnarray}
Here ${\tilde p}'(\Omega')$ is the out-going $4$-momentum vector
depending on the scattering angle $\Omega'$\,.
In deriving the second equality we use Eq.~(\ref{incoming}) and pass
to the observer's frame with ${\tilde p}_{1,2}$ standing for the
out-going $4$-momentum vector of the two particles. To proceed
further we introduce the differential cross section in the CM frame
$\sigma_{\rm cm}$ defined as $dx'dy'\equiv \sigma_{\rm
cm}(p_1,p_2\rightarrow {\tilde p}_1,\, {\tilde p}_2) d\Omega'$\,. By
definition, $\sigma_{\rm cm}$ is a manifestly covariant concept.
Moreover, in the observer's frame the quantity $u'^0 |{\bf u}'|-u'^0
(-|{\bf u}'|)$ may be written as $\sqrt {(u_1\cdot u_2)^2-1}$\,.
Collecting everything together and taking into account
Eq.~(\ref{f}), we obtain
\begin{eqnarray}
&& \left\{u^\mu_1 \partial_{\mu 1}  - \int_{\Sigma_2 \otimes U^4_2 }
d\Sigma_{\mu 2} d^4 p_2u^\mu_2\lambda \LI'_{12}f_0(2) \right\}
f_0(1) \nonumber\\
&=& \mathbb{K}_2 [f_0] \,.
\label{reducedonebody1}
\end{eqnarray}
Here the collision integral is
\begin{eqnarray}
\mathbb{K}_2[f] &=& \int\!\!\!\! \int d\Omega' \frac{d^3
p_2}{p_2^0}\, \sigma_{\rm cm}
\sqrt {(u_1\cdot
u_2)^2-1}  \label{I2result3}\\
&& \times \{f_0(x_1,{\tilde p}_1)f_0(x_1,{\tilde p}_2)
-f_0(x_1,p_1)f_0(x_1,p_2)\}. \nonumber
\end{eqnarray}
Eq.~(\ref{I2result3}) is the well known relativistic Boltzmann
collision integral \cite{Cercigani02,vanLeeuwen80}, where the
details of the two-body scattering enter through the invariant
differential cross section $\sigma_{\rm cm}$\,.

\section{Correlation at relativistic many-body equilibrium}
\label{equilibrium}

It is important that the perceptions of a macroscopic observer
correspond to the hierarchy of physical correlation functions namely
$\overrightarrow{\mathfrak{N}}_\infty$ rather than
$\overrightarrow{\mathfrak{D}}$\,,
and the hierarchy is fully determined by the solution to the general
transport equation (\ref{reducedonebody}) with the collision
integral given by Eq.~(\ref{averagekinetic10}). Obviously, the
processes with regard to the physical distribution described by
Eq.~(\ref{reducedonebody}) are observer-independent which, in the
chosen inertial frame, may be parametrized by the coordinate time.
For rarefied plasmas Eq.~(\ref{reducedonebody}) becomes a local
equation, and the collision integral generally admits the so-called
``collision invariants'' which, typically, are $\phi(p) \equiv
1,p^\mu$\,, i.e.,
\begin{equation}
\int_{U^4} d^4p \, \phi(p)\, \mathbb{K}[f(x,p)]=0 \,.
\label{invariant}
\end{equation}
They amount to the particle number and the energy-momentum
conservation law, from which macroscopic relativistic hydrodynamics
follows.

As the system reaches global or local equilibrium the physical
one-body distribution function nullifies the collision integral,
i.e., $\mathbb{K}[f_{\rm eq}]=0$\,. The former differs from the
latter in the spacetime independence of $f_{\rm eq}$\,. By inserting
$f_{\rm eq}$ into Eq.~(\ref{correlationresult2}) we obtain
\begin{widetext}
\begin{eqnarray}
\langle \Gamma_j(1,\cdots\,,j) |\overrightarrow{\mathfrak{N}}_\infty
&=& \int_0^{\infty} ds\!\!\int_0^s ds' \sum_{n=j}^\infty
\prod_{i=j+1}^n \int_{\Sigma_i \otimes U^4_i } d\Sigma_{\mu i} d^4
p_i\, u^\mu_i \, \prod_{k=1}^n f_{\rm eq}(x_k,p_k) \nonumber\\
&& \qquad \qquad \qquad \times \langle \Gamma_j (1,\cdots\,,j) | \C
{\hat {\mathfrak{U}}}^0 (s-s'){\hat {\mathfrak{E}}}(s') \V
\exp(-s\V\G\V)|1|\cdots |n\rangle
    \label{correlationequilibrium}
\end{eqnarray}
\end{widetext}
\noindent which, in principle, gives the entire hierarchy of
physical equilibrium correlation functions. Therefore, the
conjecture of Hakim on the many-body equilibrium \cite{Hakim67} is
justified. In Ref.~\cite{Hakim67} only the first component, i.e.,
$\langle 12|\overrightarrow{\mathfrak{N}}_\infty$ is formulated
using the weak coupling approximation. This hierarchy, together with
$f_{\rm eq}(x,p)$ fully captures the many-body equilibrium of
rarified relativistic plasmas. It allows one to go beyond the
kinetic approximation \cite{Israel84,Kandrup84} to formulate the
relativistic many-body equilibrium basing on the
action-at-a-distance formalism.

\subsection{J{\"u}ttner equilibrium}
\label{Juettner1}

Let us now exemplify these general principles in the case of
$\mathbb{K}_{1,2}[f]$\,, which can be easily shown to preserve
Eq.~(\ref{invariant}). This then results in the relativistic
hydrodynamics--a well established subject \cite{vanLeeuwen80} which
we shall not proceed further. Encompassed by the hydrodynamics the
system irreversibly evolves into the local J{\"u}ttner equilibrium.
Indeed, define the local entropy flux $S^\mu (x)$ as follows:
\begin{equation}
S^\mu (x) \equiv -\int d^4p\,2m \theta(p^0)\delta (p^2-m^2) u^\mu
f_0(x,p)\ln f_0(x,p) \,, \label{entropyflow}
\end{equation}
where $f_0(x,p)$ is defined by Eq.~(\ref{f}). Then, from the
collision integral (\ref{I2result3}) the $H$-theorem follows
\cite{vanLeeuwen80}, i.e.,
\begin{eqnarray}
\sigma_s(x)\equiv \partial_\mu S^\mu (x) \geq 0\,.
\label{entropyproduction}
\end{eqnarray}
The local entropy production $\sigma_s(x)$ vanishes wherever
$f_0(x,p)$ reaches the local J{\"u}ttner equilibrium:
\begin{equation}
f_J(x,p) = \frac{\rho(x)\beta(x)}{4\pi m^2 K_2 (m\beta(x))}
e^{-\beta^\mu(x) p_\mu}\,. \label{Juettner}
\end{equation}
Here $\rho(x)$ is the invariant particle number density, $K_2$ is
the modified Bessel function of order two, and $\beta^\mu(x)$ is the
(local) reciprocal temperature timelike $4$-vector defining the
(local) temperature $T(x)$ through $1/k_BT(x) \equiv \sqrt {\beta
(x)^2}$\,. Suppressing all the spacetime dependence of the
parameters of $f_J(x,p)$ further leads to the (global) J{\"u}ttner
equilibrium:
\begin{equation}
f_{\rm eq}(p)\equiv f_J(p) 2m \theta(p^0)\delta (p^2-m^2) \,.
\label{feq}
\end{equation}

The debate on the special relativistic version of the
Maxwell-Boltzmann distribution has stemmed from
Ref.~\cite{Schieve89} where a different Boltzmann equation, though
manifestly covariant, is proposed. There, proceeding along the line
similar to Eqs.~(\ref{entropyflow})-(\ref{Juettner}) results in an
alternative one-body equilibrium distribution. In the present work
by the proof of the relativistic Landau and Boltzmann equation
namely Eq.~(\ref{reducedonebody}) with the collision integral
(\ref{BBKcollisionresult}) and (\ref{I2result3}) is implied the
following important fact: In rarefied plasmas the J{\"u}ttner
distribution is no longer a phenomenological hypothesis, rather, is
well justified at the full microscopic level and, importantly, suits
the covariant principle. Thus, we provide a solid support for very
recent $1$-dimensional numerical simulation \cite{Hanggi07} and
advocates the critical analysis on alternatives to the J{\"u}ttner
distribution \cite{Debbasch08}. Notice that the spatial
dimensionality in the numerical simulation differs which, we
believe, is of minor importance. (In fact, for plasmas the
relativistic many-body dynamics and thereby the collision integral
are substantially simplified in the $1$-dimensional geometry. There,
the electromagnetic interaction is purely longitudinal.) What is
crucial is that, there, the underlying many-body dynamics is
dictated by point-like collisions and is well described by the
action-at-a-distance formalism and, as a result, the entire scope of
the present manifestly covariant correlation dynamics applies.

\subsection{Equal-time two-body correlation}
\label{correlation4}

We further proceed to calculate the correlation function at the
J{\"u}ttner equilibrium. For simplicity let us consider the simplest
two-body correlation function, which is $C_{\rm
eq}(x_1,p_1;x_2,p_2)\equiv \langle 12|
\overrightarrow{\mathfrak{N}}_\infty$\,. Here we use the subscript
``eq'' to denote that the physical one-body distribution is at
global J{\"u}ttner equilibrium, i.e., $
f(x_i,p_i)=f_{\rm eq}(p_i)$\,. In particular, to investigate the
behavior of the spacelike correlation function we will calculate
below the so-called equal-time correlation function.

\subsubsection{Bare correlation function}
\label{bare}

\begin{figure}
\begin{center}
\leavevmode \epsfxsize=8cm \epsfbox{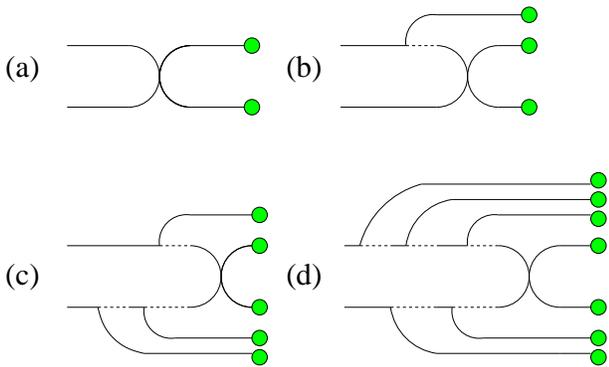}
\end{center}
\caption{Diagram of bare two-body correlation (a) and typical
diagrams leading to short-ranged correlation (b)-(d). Solid circles
stand for the J{\"u}ttner equilibrium.}
  \label{ring}
\end{figure}

To the lowest order interaction expansion [see Fig.~\ref{ring} (a)]
Eq.~(\ref{correlationequilibrium}) gives a bare two-body correlation
function:
\begin{eqnarray}
&& C^0_{\rm eq}(x_1,p_1;x_2,p_2) \nonumber\\
&=& \int_0^{\infty} ds\, \langle 12|\U^0(s)\, \lambda \Li'
|1|2\rangle \,f_{\rm eq}(p_1)f_{\rm eq}(p_2)\,.
    \label{twobodybare}
\end{eqnarray}
Under the relativistic impulse approximation (see
Appendix~\ref{impulseLiouville} for details) it is translationally
invariant, i.e., $C_{\rm eq}(x_1,p_1;x_2,p_2)\equiv C_{\rm
eq}^0(x_1-x_2,p_1,p_2)$\,. Indeed, inserting the matrix elements of
$\U_0$ and $\lambda \Li'$ into Eq.~(\ref{twobodybare}) we find
\begin{eqnarray}
C_{\rm eq}^0(x_1-x_2,p_1,p_2) = \int \frac{d^4 k}{(2\pi)^4}\, e^{ ik
\cdot (x_1-x_2)}\, {\tilde C}_{\rm eq}^0(k,p_1,p_2) \nonumber\\
\label{twobodybare2}
\end{eqnarray}
with
\begin{widetext}
\begin{eqnarray}
{\tilde C}_{\rm eq}^0(k,p_1,p_2) &=& -\frac{i 8\pi^2 e^2}{k^2}
\frac{1}{ik\cdot (p_1-p_2)} \{\delta
(k\cdot p_2) [k\cdot\beta (p_1\cdot p_2)-\beta\cdot p_2 (k\cdot p_1)] \nonumber\\
&& \qquad \qquad \qquad \qquad \quad -\delta (k\cdot p_1) [k\cdot
\beta (p_1\cdot p_2)-\beta\cdot p_1 (k\cdot p_2)] \}f_{\rm
eq}(p_1)f_{\rm eq}(p_2) \,,
    \label{twobodybare1}
\end{eqnarray}
where the wave vector $k^\mu\equiv (\omega, {\bf k})$\,.

Since the equilibrium is {\it global} we may take the advantage of
the Lorentz invariance of Eq.~(\ref{twobodybare2}) and choose the
observer's frame where $\beta^\mu =(1/k_B T,0)$\,, and the spacelike
$3$-surface is the usual $3$-dimensional Euclidean space. In this
frame the bare equal-time correlation function is
\begin{eqnarray}
C_{\rm eq}^0({\bf x}_1-{\bf x}_2,p_1,p_2) &=& \int \frac{d{\bf
k}}{(2\pi)^3}\, e^{ i{\bf k} \cdot ({\bf x}_1-{\bf x}_2)} \, {\cal
C}_{\rm
eq}^0({\bf k},p_1,p_2) \,, \label{equaltimecorrelation}\\
{\cal C}_{\rm eq}^0({\bf k},p_1,p_2) &=& \int \frac{d\omega
}{2\pi}{\tilde C}_{\rm eq}^0(\omega,{\bf k},p_1,p_2)\,. \nonumber
\end{eqnarray}
Notice that in the observer's frame Eq.~(\ref{twobodybare1}) is
written as
\begin{eqnarray}
{\tilde C}_{\rm eq}^0(\omega, {\bf k},p_1,p_2) &=& -\frac{i 8\pi^2
e^2}{k^2} \bigg\{\delta (k\cdot p_2) \bigg[\frac{\omega}{k_BT}
\frac{(p_1\cdot p_2)}{i(\omega p_1^0-{\bf k}\cdot {\bf p}_1)}
+ \frac{ip^0_2}{k_BT} \bigg]\nonumber\\
&& \quad \quad \quad \quad +\delta (k\cdot p_1)
\bigg[\frac{\omega}{k_BT} \frac{(p_1\cdot p_2)}{i(\omega p_2^0-{\bf
k}\cdot {\bf p}_2)} + \frac{ip^0_1}{k_BT}\bigg] \bigg\}f_{\rm
eq}(p_1)f_{\rm eq}(p_2) \,. \label{equaltimecorrelation1}
\end{eqnarray}
Integrating out $\omega$ we find
\begin{eqnarray}
{\cal C}_{\rm eq}^0({\bf k},p_1,p_2) \propto -
\frac{f_{\rm eq}(p_1)f_{\rm eq}(p_2)}{|{\bf k}|^2} \,,
\label{equaltime}
\end{eqnarray}
where the irrelevant numerical overall factor is ignored.
Eq.~(\ref{equaltime}) shows that the bare equal-time two-body
correlation is long-ranged, i.e., $C_{\rm eq}^0({\bf x}_1-{\bf
x}_2,p_1,p_2) \propto |{\bf x}_1-{\bf x}_2|^{-1}$ reflecting the
long-ranged nature of electromagnetic interactions.

\subsubsection{Short-ranged equal-time correlation: Collective dielectric effects}
\label{shortcorrelation}

To go beyond the weak coupling approximation we extend the so-called
{\it ring approximation} \cite{Balescu} to the relativistic context.
That is, we will sum up all the so-called {\it ring diagrams}.
[Typical diagrams are given in Fig.~\ref{ring} (b)-(d).] They are
obtained from Fig.~\ref{ring} (a) in the following way: The
propagating lines to the left of the vertex--associated with no
particle annihilation--are dressed by a sequence of vertexes. For
each of them an additional particle joins the vertex from the right
and continues propagating to the left, and eventually is annihilated
at the next vertex. (It can be shown that the propagating lines to
the right suffer {\it no} renormalization effects.) In selecting
these diagrams it is implied that in
Eq.~(\ref{correlationequilibrium}) $\exp(-s\V\G\V)$ is set to be
unity. Summing up all these diagrams we find that the two-body
correlation function solves the following Dyson equation:
\begin{eqnarray}
C_{\rm eq}(x_1,p_1;x_2,p_2) &=& C_{\rm eq}^0(x_1,p_1;x_2,p_2) +
\int_0^\infty ds\, e^{-s(u^\mu_1\partial_{\mu
1}+u^\mu_2\partial_{\mu 2})} \int_{\Sigma_3\otimes
U^4_3}d\Sigma_{\mu 3} d^4 p_3 u^\mu_3 \nonumber\\
&& \qquad \qquad \qquad \times \left\{\lambda{\hat {\cal L}}'_{13}
f_{\rm eq}(p_1)C_{\rm eq}(x_3,p_3;x_2,p_2) +\lambda{\hat {\cal
L}}'_{23} f_{\rm eq}(p_2)C_{\rm eq}(x_3,p_3;x_1,p_1)\right\} \,.
\label{ringcorrelation}
\end{eqnarray}

Let us now pass to the observer's frame chosen above, where
$t_1=t_2=t_3$ and calculate the equal-time correlation using
Eq.~(\ref{ringcorrelation}). At global equilibrium such a
correlation function does not depend on the coordinate time.
Consequently, upon passing to the spatial Fourier transformation we
obtain
\begin{eqnarray}
{\cal C}_{\rm eq}({\bf k},p_1,p_2) &=& {\cal C}^0_{\rm eq}({\bf
k},p_1,p_2)
  + i8\pi^2e^2 \int\frac{d\omega}{2\pi} \frac{1}{\omega^2-|{\bf k}|^2} \int d^4 p_3 u_3^0\frac{\delta (k\cdot
p_3)}{i{\bf k}\cdot ({\bf p}_1-{\bf p}_2)}
\nonumber\\
&&\times \bigg\{\left[\frac{\omega}{k_B T} (p_1\cdot
p_3)-\frac{p_3^0}{k_B T} (k\cdot p_1)\right]f_{\rm eq}(p_1){\cal
C}_{\rm eq}({\bf k},p_3,p_2)
 \nonumber\\
&& \,\,\,\, -\left[\frac{\omega}{k_B T} (p_2\cdot
p_3)-\frac{p_3^0}{k_B T} (k\cdot p_2)\right]f_{\rm eq}(p_2){\cal
C}_{\rm eq}({\bf k},p_3,p_1)\bigg\} \,, \label{ringcorrelation1}
\end{eqnarray}
where ${\cal C}_{\rm eq}({\bf k},p_i,p_j)$ is the Fourier
transformation of ${\cal C}_{\rm eq}({\bf x}_i-{\bf
x}_j,p_i,p_j)$\,. To further proceed we choose the direction of
${\bf k}$ as the $x$-axis and integrate out $\omega$\,. It is
natural to expect that ${\cal C}_{\rm eq}({\bf k},p_i,p_j)$
possesses the spherical symmetry with respect to ${\bf p}_{i,j}$\,.
As a result, we obtain
\begin{eqnarray}
{\cal C}_{\rm eq}({\bf k},p_1,p_2) &=& {\cal C}^0_{\rm eq}({\bf
k},p_1,p_2)
  - \frac{4\pi^2e^2}{k_BT} \frac{1}{|{\bf k}|^2}\frac{1}{p_{1x}-p_{2x}}
\bigg\{p_{1x}f_{\rm eq}(p_1)\int d^4 p_3 u_3^0\, {\cal C}_{\rm
eq}({\bf k},p_3,p_2)\nonumber\\
&& \qquad \qquad \qquad \qquad \qquad \qquad \qquad \quad \,\,
-p_{2x}f_{\rm eq}(p_2)\int d^4 p_3 u_3^0\, {\cal C}_{\rm eq}({\bf
k},p_3,p_1)\bigg\} \,. \label{ringcorrelation4}
\end{eqnarray}
\end{widetext}
Inserting Eq.~(\ref{equaltime}) into it we find the solution to be
\begin{eqnarray}
C_{\rm eq}({\bf k},p_1,p_2) \propto
- \frac{f_{\rm eq}(p_1)f_{\rm eq}(p_2)}{|{\bf k}|^2+ \lambda_{\rm
D}^{-2}} \,.
\label{ringcorrelationDebye}
\end{eqnarray}
Eq.~(\ref{ringcorrelationDebye}) with $\lambda_{\rm D}$ given by
Eq.~(\ref{ringcorrelation2}) fully agrees with the result of
Klimontovich who used a completely different (nonmanifestly
covariant) approach \cite{Klimontovich67}. They suggest that the
equaltime two-body correlation function (in the phase space)
displays an exponential decay with the correlation radius
$\lambda_{\rm D}$\,, which is none but the nonrelativistic Debye
length. The relativistic insensibility (rest mass) of $\lambda_{\rm
D}$ was noticed long time ago \cite{Landau}. Comparing
Eq.~(\ref{ringcorrelationDebye}) with Eq.~(\ref{equaltime}) we find
that despite of the appearance of the transverse electromagnetic
interaction, which is screened in a dynamical manner
\cite{Litim02,Landau8}, the collective screening of the longitudinal
electromagnetic interaction renders the equal-time correlation
function short-ranged. Moreover, the static longitudinal
permittivity is found to be
\begin{eqnarray}
\varepsilon^\parallel(\omega\rightarrow 0, {\bf k}) - 1 =
\frac{1}{(|{\bf k}|\lambda_{\rm D})^2}\,.
\label{longitudinaldielectric}
\end{eqnarray}

Thus, the ring diagrams suffice to describe the collective
dielectric effects, justifying the introduced lower cutoff
$\lambda_{\rm D}^{-1}$ in the relativistic Landau collision integral
$\mathbb{K}_1[f]$\,. Of course, to heal the infrared divergence of
$\mathbb{K}_1[f]$ accurately we need to sum up all the
nonequilibrium ring diagrams as depicted in Fig.~\ref{kineticfig}
(b). This is technically far beyond the scope of the present paper
and we leave for future studies.

\section{Conclusions}
\label{conclusion}

The present work is motivated by very recent physical and
mathematical progresses in $1$-dimensional relativistic many-body
systems \cite{Hanggi07,Kirpichev07}. Although the attempts of
reconciling the correlation dynamics and the special relativity were
undertaken by the Brussel-Austin school long time ago, the
relativistic correlation dynamics was formulated in a {\it
nonmanifestly} covariant manner \cite{Prigogine65,Balescu64}.
Because of this drawback such a theory has been less advanced so far
with regard to the practical applicability. The possibility of
formulating a manifestly covariant correlation dynamics was first
discussed by Israel and Kandrup \cite{Israel84,Kandrup84}.
Unfortunately, there have been no progresses toward this direction.
In this series of papers we substantially extend the analysis of the
relativistic classical nonequilibrium statistical mechanics of full
interacting many-body systems by Israel and Kandrup in the context
of special relativity. Then, a manifestly covariant correlation
dynamics results, which bridges macroscopic phenomena such as
hydrodynamics and equilibrium and microscopic deterministic
relativistic many-body dynamics. It is a statistical theory of
ensembles of a bundle of world lines, and is suitable for studies of
dynamics of distribution functions of full classical many-body
systems. For a summary, we carry out the following program:

(i) For a classical interacting system composed of $N$ particles the
well established action-at-a-distance formalism naturally introduces
an $8N$-dimensional $\Gamma$ phase space, for which we introduce a
probability distribution function. From the conservation law a
single-time Liouville equation follows, describing the manifestly
covariant global evolution of the distribution function of the
$\Gamma$ phase space.

(ii) For the manifestly covariant single-time Liouville equation we
perform the correlation dynamics analysis. First, we introduce the
correlation pattern representation, where the cluster expansion of
reduced distribution functions is formulated in a manifestly
covariant manner. Then, it is found that the evolution of the full
$N$-body distribution function (of the $\Gamma$ phase space) may be
reduced into that of the one-body distribution function (of the
$8$-dimensional $\mu$ phase space) which is described by a closed
nonlinear equation. However, the evolution of such a one-body
distribution function may not be observable. Rather, the stationary
solution of this closed nonlinear equation results in an exact
general transport equation. The solution of the latter is physically
observable and fully responsible for macroscopic hydrodynamics and,
very importantly, determines the entire physical correlation
functions. Such picture is encapsulated in
Eqs.~(\ref{reducedonebody}) and (\ref{correlationresult2}), which
constitute a manifestly covariant version of the Bogoliubov
functional assumption--the general principle underlying kinetic
processes of many-body systems.

(iii) The collision integral of the general transport equation is
expressed in terms of the density expansion. Therefore, for
low-density systems only the leading term needs to be taken into
account. For such obtained collision integral the perturbative
theory with respect to the interaction is formulated. Then, under
the relativistic impulse approximation we recover the Landau
collision integral by keeping the leading order interaction
expansion, while recover the Boltzmann collision integral by keeping
the entire interaction expansion.

(iv) The analysis of the entropy production due to the Landau and
the Boltzmann collision integral shows that a low-density system
(for example, rarified plasmas) tend to be driven to the (local)
J{\"u}ttner equilibrium. Replacing the physical one-body
distribution function in the hierarchy (\ref{correlationresult2})
with the (global) J{\"u}ttner equilibrium, we justify the long
standing many-body equilibrium conjecture of Hakim. In particular,
we calculate explicitly the (static) two-body correlation function
for plasmas and analyze the collective screening effects.

We remark that the emergence of a manifestly covariant formulation
of the Bogoliubov functional assumption is by no means obvious.
Indeed, in relativity the breakdown of the simultaneity as a
covariant concept profoundly affects the mathematical structure of
the correlation dynamics. Among the most important modifications is
that the microscopic states, in general, are no longer the
representation points in the $\Gamma$ phase space, rather, are a
bundle of $N$ (segments of) particle world lines. (There do exist
exceptions, for example, the case studied in
Ref.~\cite{Kirpichev07}.) Indeed, the introduction of the $\Gamma$
phase space is merely suggested by the form of the motion equations.
In particular, that the force does not depend on the acceleration of
the acted particle (Thus, the radiation reaction is ignored as
electromagnetic interactions concerned.) is crucial to the present
choice of the $\Gamma$ phase space. As a result, the building block
operator namely the two-body interacting Liouvillian and the
distribution function are determined not only by the positions and
the momenta of two participating particles, but also by their world
lines (more precisely, the phase trajectories) passing through the
given positions and momenta. The very origin of the Bogoliubov
functional assumption namely Eqs.~(\ref{reducedonebody}) and
(\ref{correlationresult2}) is that the deterministic many-body
dynamics lose the memory of the given (segments of) particle world
lines at large global proper times.

In fact, it is the nonmanifestly covariant version of the Bogoliubov
functional assumption, formulated in a rather straightforward
manner, that is commonly adopted in the literature of plasmas
\cite{Lu94,Naumov81,Tian01}. However, the proof has been lacking.
The manifestly covariant Bogoliubov functional assumption namely
Eqs.~(\ref{reducedonebody}) and (\ref{correlationresult2}) justifies
such an formulation. Indeed, it was first noticed by Dirac, Fock,
and Podolsky \cite{Dirac32} and stressed by Israel \cite{Kandrup84}
that it is legitimate to identify the coordinate times of particles,
i.e., to set $t_1=t_2=\cdots =t_N\equiv t$ for the manifestly
covariant equation $\Li {\cal N}=0$\,. In doing so we obtain
Eqs.~(\ref{reducedonebody}) and (\ref{averagekinetic10}) with all
the coordinate times set to be the same. Next, as (implicitly) done
in various nonmanifestly covariant theories we again apply the
relativistic impulse approximation. The key point is that such an
approximation does not destroy the (manifest) covariance of Lorentz
force. Then, the collision integral (\ref{kineticria4}) remains
unaffected except that there $t_1$ and $t_2$ have to be identified.
For this collision integral let us keep only the leading order
interaction expansion. Immediately, we arrive at a relativistic
Landau collision integral which is a nonmanifestly covariant version
of $\mathbb{K}_1[f]$\,. Thus, we explain the insightful observation
made in Ref.~\cite{Israel84}, where it was pointed out that to
derive relativistic transport equations by using a nonmanifestly
covariant approach may not be a problem as electromagnetic
interactions concerned.

We stress that upon passing from the motion equations
(\ref{Newton1}) and (\ref{Newton}) to the single-time Louville
equation (\ref{Liouville1}) some information of the underlying
many-body dynamics is lost. Thus, they are not equivalent. The
manifestly covariant correlation dynamics stems from the latter.
Surprisingly, for rarefied plasmas the present correlation dynamics
recovers the Landau and the Boltzmann equation which are fully
responsible for various macroscopic phenomena such as hydrodynamics
and the (local) equilibrium. Thus, we are led to the following conjecture:\\

\noindent {\it For systems with arbitrary macroscopic particle
number density the manifestly covariant single-time Louville
equation (\ref{Liouville1}) completely determines macroscopic
relativistic hydrodynamics.}\\

Although the relativistic Landau equation has been justified before
especially by the manifestly covariant nonequilibrium statistical
mechanics developed in Refs.~\cite{Israel84,Kandrup84}, compared to
the earlier theories the manifestly covariant correlation dynamics
presented here has the additional advantage in going beyond the weak
coupling approximation: First, it clearly shows that the
relativistic Landau equation is the leading order interaction
expansion of the relativistic Boltzmann equation. It is the two-body
scattering approximation that is used in deriving the latter
equation, and the collision integral (with the accuracy of the
leading order density expansion) is expressed in terms of the
interaction expansion. The present theory allows one to sum up this
expansion and, as a result, the scattering cross section is beyond
the weak coupling level. Second, the general collision integral
(\ref{averagekinetic10}) and the hierarchy of correlation functions
(\ref{correlationresult2}) further allow one to systematically go
beyond the two-body scattering approximation. This is exemplified in
the calculations of the static two-body equilibrium correlation. It
is promising in understanding the infrared divergence of the
relativistic Landau collision integral where collective dynamic
screening phenomena play crucial roles, and in generalizing the
Balescu-Lenard equation \cite{Balescu60} so as to suit relativity
principles. (To our best knowledge, so far only the nonmanifestly
covariant Balescu-Lenard equation has been obtained
\cite{Markov95,Klimontovich67,Lu94}.) Finally, the present theory
allows one to explore the anomalous transport phenomenon arising
from the long-time tail correlation effects \cite{Dorfman70}. These
issues seem to be far beyond the reach of earlier theories such as
Refs.~\cite{Israel84,Kandrup84}.

Closing this paper we remark that although this series of papers
focuses on the classical correlation dynamics, it may have important
implications on the relativistic quantum nonequilibrium statistical
mechanics. Indeed, to formulate the latter at the level of full
many-body quantum dynamics has been of long term interests and
remained controversial so far \cite{Schieve05}. Besides of the
manifest/nonmanifest covariance issue another severe difficulty
arises as the quantization of relativistic many-body classical
dynamics concerned. In order not to lose the manifest covariance
Schieve recently started from the so-called covariant Hamiltonian
dynamics and formulated a theory of relativistic quantum statistical
mechanics \cite{Schieve05}. It is important to notice that the
quantum Boltzmann equation derived there is with regard to
``events'' in the spacetime rather than particles. For this reason
neither can this equation be justified by the microscopic approach
\cite{DuBois72} basing on the generalization of traditional
nonequilibrium Green's function theories \cite{Schwinger,Kadanoff}
or by refined relativistic nonequilibrium quantum field theories
\cite{Hu88,Chou85}, nor the classical limit of this equation be
justified by other classical microscopic theories
\cite{Israel84,Kandrup84}. In contrast, the quantum transport
equations obtained by complete quantum treatment, in the classical
limit, agree with the transport equations shown in the present
paper. This justifies the legitimacy of basing classical
relativistic statistical mechanics on the action-at-a-distance
formalism. Many natural problems thereby arise: {\it May it be
possible to be extended to the quantum case?} What is the quantum
counterpart of the proper time-parametrized evolution? In
particular, over a field theory the advantage of proceeding along
this line to formulate a quantum statistical theory is to allow for
nonperturbative treatment and to encompass some global properties,
as admitted by many authors (for example, see Ref.~\cite{Hu88}).
Unfortunately, to solve these problems is by no means an easy task
because of the lack of Hamiltonian of the action-at-a-distance
formalism. A possible prescription is to enlarge the configuration
space to accommodate the field degrees of freedom. Treating the
particles and the fields on the same footing one may proceed to
construct a many-body Green function by adopting the prescription of
Feynman \cite{Feynman50,Fainberg95}, namely to express it in terms
of the functional integral over all the paths of both particles and
fields parametrized by separate proper times. Furthermore, it is
expected that such a many-body Green function satisfies a
St{\"u}ckelberg-type equation \cite{Stueckelberg} suitable for the
use of the Wigner function technique. As such the entire theoretical
scope of nonrelativistic quantum correlation dynamics \cite{Balescu}
might become applicable. We leave the detailed analysis for future
studies.

\acknowledgements

I am deeply grateful to Q. K. Lu for numerous fruitful discussions
at the early stage of this work, and especially to S. L. Tian for
invaluable help. I also would like to thank M. Courbage, J. R.
Dorfman and M. Garst for useful conversations, and especially to C.
Kiefer for his interests and encouragements. This work is supported
by Transregio SFB 12 of the Deutsche Forschungsgemeinschaft and was
partly done in Institute of Henri Poincare.

\begin{appendix}

\section{Relativistic impulse approximation for electromagnetic interactions}
\label{EM}

In this appendix we wish to study the relativistic impulse
approximation to two-body interacting Liouvillian for classical
electromagnetic interactions.

\subsection{Exact two-body interacting Liouvillian}
\label{exactLiouville}

We notice that, in general, the interacting force $F_{ij}^\mu$ has
the form as follows \cite{Tian09,Martinez06}:
\begin{eqnarray}
F_{ij}^\mu(x_i,p_i) = \int_{-\infty}^{+\infty}d\tau_j\, s
(\rho_{ij})\, {\cal F}^{\mu \nu}
|_{x_j=x_j(\tau_j)} \, p_{\nu i} \,, \label{force}\\
\rho_{ij} = (x_{\mu i}-x_{\mu j}) (x^\mu_i-x^\mu_j)\,, \qquad \qquad
\qquad \nonumber
\end{eqnarray}
where ${\cal F}^{\mu \nu}$ is an antisymmetric tensor, and the role
of function $g (\rho_{ij})$ is to invariantly connect $x_i$ with one
(or several) points at the world line $x_j(\tau_j)$\,. Let us
exemplify Eq.~(\ref{force}) and derive the exact form of two-body
interacting Liouvillian in classical electrodynamics. For this
purpose we employ the Wheeler-Feynman formalism \cite{Fokker29} and
start from the Fokker action which crucially allows us to eliminate
the self-action:
\begin{widetext}
\begin{eqnarray}
S = -m \sum_i \int ds_i ({\bar u}_{\mu i} {\bar u}^\mu_i)^{1/2} -
e^2 \sum_{i<j} \int\!\!\!\!\int ds_i ds_j \, {\bar u}_{\mu i} {\bar
u}^\mu_j\, \delta (\rho_{ij}) \label{Fokker}
\end{eqnarray}
with ${\bar u}^\mu(s)\equiv dx^\mu(s)/ds$\,.

Then, we apply the variation principle to this action. Demanding
\begin{eqnarray}
\delta S = m \sum_i \int ds_i \frac{d}{ds_i}\frac{{\bar
u}_i^\mu}{\sqrt {{\bar u}_{\nu i} {\bar u}^\nu_i}} \, \delta x_{\mu
i} + e^2 \sum_{i\neq j} \int\!\!\!\!\int ds_i ds_j \,
\left\{\partial^\nu_i [{\bar u}^\mu_j\, \delta (\rho_{ij})] -
\partial^\mu_i [{\bar u}^\nu_j\, \delta (\rho_{ij})]\right\}{\bar u}_{\nu i}
\delta x_{\mu i} \equiv 0 \label{variation}
\end{eqnarray}
\end{widetext}
and defining $d\tau_i = \sqrt {{\bar u}_{\nu i} {\bar u}^\nu_i} \,
ds_i$ and $u^\mu_i\equiv {\bar u}^\mu_i/\sqrt {{\bar u}_{\nu i}
{\bar u}^\nu_i}$\,, we obtain
\begin{eqnarray}
\frac{dp^\mu_i}{d\tau_i} &=& e \sum_{j\neq i} \, F_{ij}^{\mu\nu} (x_i) u_{\nu i} \nonumber\\
F_{ij}^{\mu \nu}(x_i) &=& [\partial^\mu A^\nu_j(x) - \partial^\nu
A^\mu_j(x)]|_{x=x_i}\,, \label{EM1}
\end{eqnarray}
where $F^{\mu\nu}_{ij}$ is the electromagnetic tensor adjunct to
particle $j$\,, and all the field point is set to be the position of
particle $i$ at $\tau_i$\,, i.e., $x=x_i(\tau_i)$\,. The vector
potential, adjunct to particle $j$\,, is given by
\begin{equation}
A^\mu_j(x) = e\int \delta((x-x_j)^2)\, u^\mu_j \, d\tau_j\,.
\label{vectorpotential}
\end{equation}
Notice that here both $x_j$ and $u_j$ depend on the proper time
$\tau_j$\,. Eqs.~(\ref{EM1}) and (\ref{vectorpotential}) describe
the classical electrodynamics of electronic systems in terms of
direct interparticle action. Notice that the vector potential given
by Eq.~(\ref{vectorpotential}) satisfies Lorentz gauge:
\begin{equation}
\partial_\mu A^\mu_j(x) = 0 \,. \label{Lorentz}
\end{equation}

Inserting Eq.~(\ref{vectorpotential}) into Eq.~(\ref{EM1}) gives
Eq.~(\ref{force}) with
\begin{eqnarray}
{\cal F}^{\mu\nu} = - \frac{e^2}{m^2}\, \left[(x^\mu_i-x^\mu_j)
p^\nu_j-p^\mu_j
(x^\nu_i-x^\nu_j)\right] \,, \label{forceEM1}\\
s(\rho_{ij}) = \delta'(\rho_{ij}) \,. \qquad \qquad
 \qquad \label{forceEM2}
\end{eqnarray}
Here $\delta'(x)\equiv d\delta(x)/dx$\,.

With the help of Eqs.~(\ref{EM1}) and (\ref{vectorpotential}) the
exact two-body interacting Liouvillian is written as
\begin{widetext}
\begin{eqnarray}
\lambda \LI'_{ij} (x_i,p_i;x_j,p_j) &=& - e^2 \int d\tau \, \bigg \{
u_{\nu i} [u^\nu_{j} (\tau) \partial_i^\mu - u^\mu_{j} (\tau)
\partial_i^\nu]\, \delta ((x_i-x_j(\tau))^2) \, \frac{\partial}{\partial
p_i^\mu} \nonumber\\
&& \qquad \qquad \,\,\, + u_{\nu j} [u^\nu_{i} (\tau) \partial_j^\mu
- u^\mu_{i} (\tau)
\partial_j^\nu]\, \delta ((x_j-x_i(\tau))^2) \, \frac{\partial}{\partial
p_j^\mu} \bigg\} \,. \label{twobodyLiouville}
\end{eqnarray}

\subsection{Two-body interacting Liouvillian: relativistic impulse approximation}
\label{impulseLiouville}

Now we proceed to find the relativistic impulse approximation to the
two-body interacting Liouvillian. For this purpose we use the Dirac
identity: $\partial_\mu \partial^\mu \delta (x_\mu x^\mu)=4\pi
\delta^{(4)}(x)$ to obtain
\begin{equation}
\delta (x_\mu x^\mu) = - 4\pi \int \frac{d^4 k}{(2\pi)^4}\,
\frac{e^{ik \cdot x}}{k^2} \,. \label{EMfield}
\end{equation}
Substituting it into Eq.~(\ref{twobodyLiouville}) and integrating
out $\tau$\,, with Eq.~(\ref{impulse}) taken into account we obtain
\begin{eqnarray}
F_{ij}^\mu (x_i,p_i) = - \frac{i 8\pi^2 e^2}{m} \, \int \frac{d^4
k}{(2\pi)^4}\, \frac{e^{ ik \cdot (x_i-x_j)} }{k^2}\, \delta (k\cdot
p_j) [k^\mu (p_i\cdot p_j)-p_j^\mu (k\cdot p_i)]
\label{impulseforce}
\end{eqnarray}
\end{widetext}
and likewise for $F_{ji}^\mu (x_j,p_j)$\,. Eventually we find the
two-body interacting Liouvillian, under the relativistic impulse
approximation to be Eq.~(\ref{forceEMresult}). (Notice that there we
need to make the replacement: $1\rightarrow i\,, 2\rightarrow j$ for
the subscript.) It is important to notice that the mass-shell
constraint is preserved by the forces:
\begin{equation}
F_{ij}\cdot p_i = F_{ji}\cdot p_j=0\,. \label{impulseconstraint}
\end{equation}

\subsection{Two-body scattering in the CM frame}
\label{thirdlaw}

Suppose that the coordinate times of particle $i$ and $j$ are
identified, i.e., $t_i=t_j\equiv t$\,. Then, in order for
Eq.~(\ref{Z}) to accommodate a solution in the CM frame we need to
establish the following ($\gamma_{i,j}=\sqrt{m^2 + |{\bf
p}_{i,j}|^2}/m$)\\
\\
\noindent Lemma. {\it For the relativistic two-body dynamics
described by the following equations:
\begin{eqnarray}
&& \gamma_i \frac{d{\bf x}_i}{dt}  = \frac{{\bf p}_i}{m} \,, \qquad
\gamma_i \frac{d{\bf p}_i}{dt}  = {\bf F}_{ij}
\label{motionequation1}\\
&& \gamma_j \frac{d{\bf x}_j}{dt}  = \frac{{\bf p}_j}{m} \,, \qquad
\gamma_j \frac{d{\bf p}_j}{dt}  = {\bf F}_{ji}
\label{motionequation2}
\end{eqnarray}
with the force given by Eq.~(\ref{impulseforce}), if at $t=0$ exists
${\bf p}_i(0)+{\bf p}_j(0)=0$\,, then it holds for all $t>0$\,,
i.e., ${\bf p}_i(t)+{\bf
p}_j(t)=0$\,.}\\

{\it Proof.} This is quite obvious. First of all, enforcing
$t_i=t_j\equiv t$ we find, from Eq.~(\ref{impulseforce}), that the
force arising from interactions depends on ${\bf x}_i-{\bf x}_j$ and
${\bf p}_{i,j}$\,. Then, Eqs.~(\ref{motionequation1}) and
(\ref{motionequation2}) uniquely determine the phase trajectories:
${\bf x}_{i,j}(t)
$ provided that the initial condition: ${\bf x}_{i,j}(0)\,, {\bf
p}_{i,j}(0)$ is set.

On the other hand, consider the trajectories satisfying ${\bf
x}_i(t)-{\bf x}_0 = -({\bf x}_j(t)-{\bf x}_0) \equiv {\bf x}(t)$\,,
where ${\bf x}_0 \equiv ({\bf x}_i(0)+{\bf x}_j(0))/2 $\,. We find
that they solve Eqs.~(\ref{motionequation1}) and
(\ref{motionequation2}) by noticing
\begin{eqnarray}
{\bf F}_{ij} &=& \frac{8\pi^2 e^2}{m^2}\int_{|{\bf k}|\gtrsim
\lambda_{\rm D}^{-1}} \frac{d^3 {\bf k}}{(2\pi)^4} \frac{\sin (2{\bf
k}\cdot {\bf x})}{({\bf k}\cdot {\bf
p})^2-|{\bf k}|^2 E^2}\nonumber\\
&& \qquad \times E [{\bf k} (E^2 + |{\bf p}|^2)-2 ({\bf k}\cdot {\bf
p}) {\bf p}] \,, \label{impulseforce2}
\end{eqnarray}
where ${\bf p}(t) \equiv {\bf p}_i(t)=-{\bf p}_j(t)$ satisfying
${\bf p}=\gamma m\dot{{\bf x}}$ ($\gamma=\sqrt{m^2 + |{\bf
p}|^2}/m$), and $E\equiv \sqrt{m^2 + |{\bf p}|^2}$\,. Thus,
Eqs.~(\ref{motionequation1}) and (\ref{motionequation2}) do admit
the solution satisfying ${\bf p}_i(t)+{\bf p}_j(t)=0$ irrespective
of ${\bf x}_{i,j}(0)$\,. With the uniqueness taken into account the
lemma follows. Q.E.D.\\

{\it Remark.} The lemma shows that under the relativistic impulse
approximation the two-body scattering may be described in the usual
CM frame where ${\bf p}_i(t)+{\bf p}_j(t)=0$ and ${\bf x}_i(t)+{\bf
x}_j(t)=0$\,. Importantly, in this frame the two-body dynamics may
be reduced into the dynamics of single particle subject to the
external force given by Eq.~(\ref{impulseforce2}).

\section{Relativistic Landau collision integral}
\label{Landau}

We now calculate Eq.~(\ref{collision1}) under the relativistic
impulse approximation. For this purpose we insert the two-body
interacting Liouvillian namely Eq.~(\ref{forceEMresult}) into it.
Notice that $\lambda \LI'_{ij} (x_i,p_i;x_j,p_j)$ depends on
$x_{i,j}$ through $x_i-x_j$\,. Then,
\begin{widetext}
\begin{eqnarray}
\mathbb{K}_1[f] &=& - \left (\frac{8\pi^2 e^2}{m}\right)^2
\int_0^\infty \!\! ds \!\! \int_{\Sigma_2 \otimes U^4_2 }
d\Sigma_{\mu 2} d^4 p_2\, u^\mu_2
\int\!\!\!\! \int \frac{d^4k}{(2\pi)^4} \frac{d^4k'}{(2\pi)^4} \nonumber\\
&& \times e^{ik\cdot (x_1-x_2)}\, {\hat {\cal G}}_{12}(k)\,
e^{ik'\cdot \{x_1-x_2-s(u_1-u_2)\}}\, {\hat {\cal
G}}_{12}(k')\, f(x_1,p_2) f(x_1,p_1) \nonumber\\
&=& - \pi \left (\frac{8\pi^2 e^2}{m}\right)^2 \, \int_{\Sigma_2
\otimes U^4_2 } d\Sigma_{\mu 2} d^4 p_2\, p^\mu_2 \int\!\!\!\! \int
\frac{d^4k}{(2\pi)^4}
\frac{d^4k'}{(2\pi)^4} \nonumber\\
&& \times  e^{ik\cdot (x_1-x_2)}\, {\hat {\cal G}}_{12}(k)\,
e^{ik'\cdot (x_1-x_2)}\, \delta (k'\cdot p_1-k'\cdot p_2)\, {\hat
{\cal G}}_{12}(k')\, f(x_1,p_2) f(x_1,p_1)\,, \label{collision2}
\end{eqnarray}
where in the derivation of the second equality we have integrated
out $s$\,. Inserting the expression of ${\hat {\cal G}}_{12}(k')$
[see Eq.~(\ref{forceEMresult})] into it we obtain
\begin{eqnarray}
\mathbb{K}_1[f] &=& - \pi \left (\frac{8\pi^2 e^2}{m}\right)^2 \,
\int_{\Sigma_2 \otimes U^4_2 } d\Sigma_{\mu 2} d^4 p_2\, p^\mu_2
\int\!\!\!\! \int \frac{d^4k}{(2\pi)^4}
\frac{d^4k'}{(2\pi)^4} \nonumber\\
&& \times  e^{ik\cdot (x_1-x_2)}\, {\hat {\cal G}}_{12}(k)\,
e^{ik'\cdot (x_1-x_2)}\, \delta (k'\cdot p_1)\, \delta(k'\cdot
p_2)\, (p_1\cdot p_2)\,
\frac{k'^\mu}{k'^2}\left(\frac{\partial}{\partial
p_1^\mu}-\frac{\partial}{\partial p_2^\mu}\right)\, f(x_1,p_2)
f(x_1,p_1)\,. \label{collision3}
\end{eqnarray}
Let us carry out the spatial integral first. For this purpose we
enjoy the manifest covariance and choose $\Sigma_2$ to be the usual
$3$-dimensional Euclidean space. As a result, we obtain
\begin{eqnarray}
\int_{\Sigma_2} d\Sigma_{\mu 2} e^{i(k+k')\cdot (x_1-x_2)} =
(2\pi)^3 \delta^{(3)}({\bf k}+{\bf k}')\, e^{i(\omega+\omega')
(t_1-t_2)} \,.
    \label{spaceintegral}
\end{eqnarray}
where we keep in mind that the wave vector $k^\mu\equiv (\omega,
{\bf k})$\,. Then,
\begin{eqnarray}
&& \int_{\Sigma_2 } d\Sigma_{\mu 2} \, p^\mu_1\, e^{i(k+k')\cdot
(x_1-x_2)} \, \delta (k\cdot p_1)\, \delta (k'\cdot p_1)\,
\delta(k'\cdot p_2)
\nonumber\\
&=& \int_{\Sigma_2} d\Sigma_{\mu 2} \, p^\mu_2\, e^{i(k+k')\cdot
(x_1-x_2)} \, \delta (k\cdot p_2)\, \delta (k'\cdot p_1)\,
\delta(k'\cdot p_2)
= (2\pi)^3 \, \delta^{(4)}(k+k')\, \delta (k\cdot
p_1)\,\delta(k\cdot p_2)\,, \label{Dirac}\\
&& \int_{\Sigma_2 } d\Sigma_{\mu 2} \, k^\mu\, e^{i(k+k')\cdot
(x_1-x_2)} \, \delta (k\cdot p_1)\, \delta (k'\cdot p_1)\,
\delta(k'\cdot p_2) = 0 \,. \label{Dirac1}
\end{eqnarray}
Let us substitute the expression of ${\hat {\cal G}}_{12}(k)$ into
Eq.~(\ref{collision3}) and take Eqs.~(\ref{Dirac}) and
(\ref{Dirac1}) into account. Noticing
\begin{eqnarray}
[\delta (k\cdot p_2) k^\mu (p_1\cdot p_2),
\partial/\partial p^\mu_1] = [\delta (k\cdot p_1) k^\mu (p_1\cdot p_2),
\partial/\partial p^\mu_2]
= 0\,,
\label{commutation}
\end{eqnarray}
we arrive at
Eq.~(\ref{BBKcollisionresult}).

\section{Divergence of tensor $\epsilon^{\mu\nu}$}
\label{tensor1}

Observing Eq.~(\ref{tensor2}) it is easy to see that
$\epsilon^{\mu\nu}$ possesses the following structure
\begin{eqnarray}
\epsilon^{\mu\nu} = c_1 g^{\mu\nu} + c_2 (u_1^\mu u_1^\nu+u_2^\mu
u_2^\nu)+ c_3 (u_1^\mu u_2^\nu+u_2^\mu u_1^\nu)
\label{tensor3}
\end{eqnarray}
enforced by the Lorentz invariance, with $c_{1,2,3}$ the
coefficients which are invariant under the Lorentz transformation.
Then, from the identity: $p_{1\mu} \epsilon^{\mu\nu}=p_{2\mu}
\epsilon^{\mu\nu}=0$ we find $c_1=[(u_1\cdot u_2)^2-1] c_2\,,
c_3=-(u_1\cdot u_2)c_2$\,. Substituting these two relations into
Eq.~(\ref{tensor3}) gives
\begin{eqnarray}
\epsilon^{\mu\nu} = c_2 \{[(u_1\cdot u_2)^2-1] g^{\mu\nu} + (u_1^\mu
u_1^\nu+u_2^\mu u_2^\nu) -(u_1\cdot u_2)(u_1^\mu u_2^\nu+u_2^\mu
u_1^\nu)\} \,, \label{tensor4}
\end{eqnarray}
where $c_2$ is determined by
\begin{eqnarray}
c_2= \frac{1}{2} [(u_1\cdot u_2)^2-1]^{-1}
g_{\mu\nu}\epsilon^{\mu\nu} \,.
\label{tensor5}
\end{eqnarray}

We now come to calculate $c_2$\,. For this purpose we insert
Eq.~(\ref{tensor2}) into Eq.~(\ref{tensor5}) arriving at
\begin{eqnarray}
c_2 = e^4\, \frac{(u_1\cdot u_2)^2}{(u_1\cdot u_2)^2-1} \int d^4 k
\delta(k\cdot u_1)\delta(k\cdot u_2) \frac{1}{k\cdot k} \,.
\label{tensor6}
\end{eqnarray}
To proceed further we enjoy the Lorentz invariance and choose the
direction of ${\bf u}_1$ as the $x$-axis of the $3$-dimensional
Euclidean space. With $\omega$ integrated out we find
\begin{eqnarray}
c_2 &=& e^4\, \frac{(u_1\cdot u_2)^2}{\gamma_1\gamma_2[(u_1\cdot
u_2)^2-1]} \int d^3 {\bf k} \frac{\delta(k_x(v_{1}-v_{2x})-k_y
v_{2y}-k_z v_{2z})}{(k_x v_{1})^2-{\bf k}^2}
\nonumber\\
&=& -e^4\, \frac{(p_1\cdot p_2)^2}{\gamma_1\gamma_2[(u_1\cdot
u_2)^2-1]}\frac{1}{|v_{1}-v_{2x}|} \int
d^2k_\perp\frac{1}{\left[\frac{{\bf k}_\perp\cdot {\bf
v}_{2\perp}}{\gamma_1 (v_1-v_{2x})}\right]^2+{\bf k}_\perp^2} \,,
\label{tensor7}
\end{eqnarray}
where in the last line we use the notation: ${\bf
k}_\perp=(k_y,k_z)\,, {\bf v}_{2\perp}=(v_{2y},v_{2z})$\,.

The integral in the last line may be easily performed by passing to
the polar coordinate system, i.e., $(k_\perp\,,\theta)$ with the
axis chosen to be the direction of ${\bf v}_{2\perp}$\,. As such the
integral is factorized into the integral over $\theta$ which is
finite \cite{Grashteyn}, and the integral over $k_\perp$ which
suffers from both the ultraviolet and the infrared divergence. More
precisely,
\begin{eqnarray}
\int d^2k_\perp\frac{1}{\left[\frac{{\bf k}_\perp\cdot {\bf
v}_{2\perp}}{\gamma_1 (v_1-v_{2x})}\right]^2+{\bf k}_\perp^2} &=&
\int \frac{dk_\perp}{k_\perp} \!\! \int_0^{2\pi} d\theta\,
\frac{1}{\left(1+\frac{{\bf v}^2_{2\perp}}{\gamma^2_1
(v_1-v_{2x})^2}\cos^2\theta\right)} \nonumber\\
&=& \frac{2\pi}{\sqrt{1+\frac{{\bf v}^2_{2\perp}}{\gamma^2_1
(v_1-v_{2x})^2}}} \int \frac{dk_\perp}{k_\perp} \,.
\label{kintegral}
\end{eqnarray}
We see that, indeed, a logarithmic divergence results.
Substituting it into Eq.~(\ref{tensor7}) gives
\begin{eqnarray}
c_2 &=& -2\pi e^4\,\int \frac{dk_\perp}{k_\perp}\, \frac{(u_1\cdot
u_2)^2}{\gamma_1\gamma_2[(u_1\cdot
u_2)^2-1]}\frac{1}{\sqrt{(v_{1}-v_{2x})^2+{\bf v}_{2\perp}^2
(1-v_1^2)}} \nonumber\\
&=& -2\pi e^4\, \int \frac{dk_\perp}{k_\perp}\, \frac{(u_1\cdot
u_2)^2}{\gamma_1\gamma_2[(u_1\cdot
u_2)^2-1]}\frac{1}{\sqrt{-(1-v_1^2)(1-v_2^2)+(1-v_1v_{2x})^2}} \,.
\label{tensor8}
\end{eqnarray}
\end{widetext}
Taking into account the identity: $u_1\cdot u_2=\gamma_1
\gamma_2(1-v_1v_{2x})$ eventually we find
\begin{eqnarray}
c_2 = -2\pi e^4\, \int \frac{dk_\perp}{k_\perp}\, \frac{(u_1\cdot
u_2)^2}{[(u_1\cdot u_2)^2-1]^{3/2}}\,. \label{tensor8}
\end{eqnarray}
Inserting it into Eq.~(\ref{tensor4}) we arrive at
Eq.~(\ref{tensor}).

\end{appendix}



\begin{thebibliography}{}
\bibitem{kinetic} In order not to make any confusion we here distinguish
such a general kinetic equation from the kinetic equation of
Boltzmann and the Boltzmann-like equation in both nonrelativistic
and relativistic physics. The former, derived from the many-body
Liouville equation, is exact, closed, and highly nonlinear. It is
satisfied by the reduced one-body distribution function. The kinetic
equation of Boltzmann is a special case of this general kinetic
equation under appropriate approximations
\cite{Prigogine63,Balescu}. Notice that in many condensed matter and
high-energy literatures (for examples,
Refs.~\cite{Litim02,Chou85,Kadanoff}) the ``Boltzmann'' equation is
in fact ``Boltzmann-like'' which, typically, is the kinetic
(transport) equation of Vlasov, Fokker-Planck, Landau, or of
Boltzmann but with the scattering cross section calculated at the
level of the weak coupling (Born) approximation.
\end{thebibliography}
\end{document}